\documentclass[11pt]{article}

\usepackage{amsmath}
\usepackage{amsthm}
\usepackage{amssymb}
\usepackage{amsfonts}
\usepackage[latin1]{inputenc}
\usepackage{multirow}

 \usepackage{hyperref}
 
\usepackage{hyperref}
\hypersetup{
   colorlinks   =  true,
    linkcolor    = blue,
    citecolor    = blue,
     urlcolor	=blue,     
 }

\usepackage[usenames,dvipsnames]{color}

\newcommand{\subsect}[1]{\subsection{#1}}
\newcommand{\subsubsect}[1]{\subsubsection{#1}}

\def\be{\begin{equation}}
\def\ee{\end{equation}}
\def\bea{\begin{eqnarray}}
\def\eea{\end{eqnarray}}

   \newcommand\dd{{\rm d}}
   
   \newcommand\s{s}
    \newcommand\kk{\omega}

\newcommand\so{\mathfrak{so}_\ka^{(3+1)}}
\newcommand\SO{\rm{SO}_\ka^{(3+1)}}

\DeclareMathOperator\spn{span}

\newcommand\ka{\Lambda}

 \newcommand\NH{\rm{N}^{(3+1)}_\ka}

\newcommand\qq{p}
\newcommand{\ttqq}{ {\tilde \qq}}
\newcommand{\bttqq}{ \mathbf{\tilde \qq}}

   \def\>#1{{\mathbf#1}}                 

\def\1{\'{\i}}

  \newcommand\ca{\mathfrak{c}_\ka^{(3+1)}}

\newcommand\nh{\mathfrak{n}^{(3+1)}_\ka}

\begin{document}

\begin{center}
\ 

\bigskip

{\LARGE 
{Lorentzian Snyder spacetimes and their Galilei and Carroll limits \\
\vskip0.1cm
 from projective geometry}}
\bigskip
\bigskip

{\sc Angel Ballesteros, Giulia Gubitosi and Francisco J. Herranz}

{Departamento de F\'isica, Universidad de Burgos, 
09001 Burgos, Spain}

e-mail: {\href{mailto:angelb@ubu.es}{angelb@ubu.es}, \href{mailto:giulia.gubitosi@gmail.com}{giulia.gubitosi@gmail.com}, \href{mailto:fjherranz@ubu.es}{fjherranz@ubu.es}}

\end{center}

\begin{abstract}
We show that  the Lorentzian Snyder  models, together with their Galilei and Carroll limiting cases, can be rigorously constructed  through the projective geometry description of Lorentzian, Galilean and Carrollian spaces with nonvanishing constant curvature. The projective coordinates of such curved spaces take the role of momenta, while  translation generators over the same spaces are identified with noncommutative spacetime coordinates. In this way, one obtains a deformed phase space algebra, which fully characterizes  the Snyder model and is invariant under boosts and rotations of the relevant kinematical symmetries.
While the momentum space of the Lorentzian Snyder models is given by certain projective coordinates on (Anti-)de Sitter spaces, we discover that the  momentum space of the Galilean (Carrollian) Snyder models is given by certain projective coordinates on curved Carroll (Newton--Hooke) spaces. This  exchange between the Galilei and Carroll limits emerging in the transition from the geometric picture to the phase space picture is traced back to an interchange of the role of coordinates and translation operators.
As a physically relevant feature, we find that in Galilean Snyder spacetimes the time coordinate does not commute with space coordinates, in contrast with previous proposals for non-relativistic Snyder models, which assume that time and space decouple in the non-relativistic limit $c\to \infty$. This remnant mixing between space and time in the non-relativistic limit is a quite general Planck-scale effect found in several quantum spacetime models.
\end{abstract}

\tableofcontents


\section{Introduction}

In 1947 Hartland Snyder developed the first concrete model of a quantum noncommutative spacetime \cite{Snyder1947}, motivated by the search for a resolution of the divergencies problem in quantum field theory, which was later solved by renormalization.  In order to minimize departures from the standard theory, the model was constructed so to retain Lorentz invariance despite the discrete spectrum of the spatial coordinates operators. The kind of spacetime noncommutativity envisaged in Snyder's model implies  a deformation of the phase space algebra: while momentum operators are  commutative and have the usual transformation properties under Lorentz transformations, their commutators with coordinates are a deformation of the usual Heisenberg--Weyl commutators. This feature contributed to the revival  of the model in the '90s, after decades of near oblivion. In fact, it was realised that Snyder's work provided a neat example of the emergence of a minimal length uncertainty and the associated generalised uncertainty principle (GUP) at the Planck scale, a possibility on which developments in quantum gravity research focused the attention at that time \cite{Amati:1988tn, Konishi:1989wk, Garay:1994en, Kempf:1994su}. 

While  GUP models can be expressed in a covariant form \cite{Quesne:2006is},  that  contains the Snyder model as a special case, most applications of the GUP involve physical frameworks where only (modified) quantum mechanics is relevant, rather than relativistic quantum field theory \cite{Das:2008kaa, Pikovski:2011zk, Chang:2001bm, Nozari:2015iba}.\footnote{For recent advancements on the development of quantum field theory for the Snyder model see \cite{Meljanac:2017jyk, Mignemi:2019btr, Girelli:2010wi}.}  For this reason  GUP models are mostly studied by restricting the attention to the spatial part of phase space, assuming that time and energy retain the properties they have in usual quantum mechanics. Likewise, in physical applications the original Snyder's model is usually restricted to the Euclidean space, assuming that this, together with trivial time and energy variables, describes the appropriate quantum-mechanical ``non-relativistic" limit \cite{Mignemi:2011gr, Lu:2011it, Ivetic:2015cwa}. 

However, this seems to be an oversimplification since in the quantum-mechanical regime spacetime symmetries are described by the Galilean transformations, which can be obtained as a $c\rightarrow \infty$ limit of the Lorentz transformations.  In ordinary physics the Galilean limit does imply a separation between space-like and time-like sectors of the phase space, but this is not necessarily the case within a quantum gravity theory.  
Indeed, in this paper we show that relics of the phase space deformations that characterize the Lorentz-invariant Snyder model induce a residual noncommutativity between the spatial and time sections of the phase space after the Galilean limit is performed. This Galilean Snyder model is the main result of this work, and in order to construct it, we use the framework of projective geometry to define the deformed phase space algebra, a method which has already proven suitable for the covariant Snyder model, as we review in detail in this paper and briefly summarize in the following.

The essential idea behind the Snyder model is to take as noncommutative position operators the generators of infinitesimal translations on a maximally symmetric curved space (these translations are  noncommutative since the manifold is curved), and to identify the momentum space with the very same (commutative) curved space, but described in terms of projective Beltrami coordinates. In particular, the curved space originally chosen by Snyder is the (3+1)-dimensional de Sitter space $\mbox{dS}^{(3+1)}$, constructed as the homogeneous space
\be
\mbox{dS}^{(3+1)}=G/H={\rm SO}(4,1)/{\rm SO}(3,1),
\ee
where ${\rm SO}(4,1)$ is the group of isometries, and  the isotropy subgroup $H={\rm SO}(3,1)$ is the Lorentz group. In this way, the position operators identified with the noncommutative translations on the homogeneous space are invariant under the isotropy subgroup (Lorentz) by construction, and the momentum space, defined in terms of the $\mbox{dS}^{(3+1)}$ projective Beltrami coordinates, is also Lorentz invariant.

The aim of this paper is to show that such projective geometry approach enables the construction of other ``Snyder-type" noncommutative spacetime models and their associated phase space algebras in a ``canonical" way induced by the projective geometry of kinematical groups and spaces. This can be done by starting from a generic homogeneous  space $M=G/H$ with constant curvature. Then the Snyder-type noncommutative spacetime  is defined by identifying its coordinates with the infinitesimal generators of translations onto $M$, while the associated curved momentum space is given by the same $M$, with momenta identified with some projective coordinates of $M$. By construction, both the noncommutative space and the curved momentum space are invariant under $H$, and the model will be characterized algebraically by the corresponding phase space algebra.

In this paper we explicitly construct the following generalized Snyder models:
\begin{itemize}
\item The Snyder spacetime and phase space algebra arising from the Anti-de Sitter space. This model, together with the original Snyder model based on de Sitter space, gives rise to what we call the Snyder--Lorentzian models, $SL^\pm$ in short, that are invariant under the Lorentz group $H={\rm SO}(3,1)$. Here, the $+$ sign refers to the original  Snyder model (de Sitter momentum space with positive cosmological constant), while $SL^-$ denotes the model with Anti-de Sitter momentum space, known in the literature as the Anti-Snyder model~\cite{Mignemi:2011gr}.

\item The Snyder--Euclidean models $SE^\pm$, arising from spherical ($+$) and hyperbolic ($-$) curved momentum spaces, which are rotationally invariant (with $H={\rm SO}(N)$ for the $N$-dimensional Snyder--Euclidean model).

\item The  Snyder--Galilean models $SG^\pm$, that can be obtained as the non-relativistic $c\to \infty$ limit of the  Snyder--Lorentzian models, and whose curved momentum spaces turn out to be the constant curvature analogues of the Carroll spacetime. These models are invariant under the non-relativistic limit of the Lorentz group, which is given by three rotations and three (commuting) boosts.

\item The Snyder--Carrollian models $SC^\pm$, that can be obtained as a second non-relativistic limit \cite{LevyLeblondCarroll}
 of the Snyder--Lorentzian models, and whose curved momentum spaces are shown to be the constant curvature analogues of the Galilean spacetime ({i.e.}, the Newton--Hooke spacetimes).  Such a  non-relativistic limit is often called in the literature    the ``ultra-relativistic" $c\to 0$ limit (see~\cite{Bergshoeff:2014jla,Hartong:2015xda,Cardona:2016ytk,Bergshoeff:2017btm,Gomis:2019} and references therein), so that we shall keep this terminology in order to distinguish it from the usual Galilean non-relativistic $c\to \infty$ limit.
 Thus  these models are  invariant under the ``ultra-relativistic" limit of the Lorentz group, which is given again by three rotations and three (commuting) boosts.\footnote{While the algebra of rotations and boosts is the same (isomorphic) for the $c\to \infty$ and $c\to 0$ limits of the Lorentz generators, the action of boosts on spacetime coordinates of course differs in the two cases, so that one can indeed distinguish between Galilean and Carrollian boosts.}

\end{itemize}

Note the  apparent exchange of the roles played by curved  Carrollian and Galilean geometries as momentum spaces for, respectively,  the new  non-relativistic and ``ultra-relativistic'' Snyder models. This will be commented upon later in the paper.

The plan of the paper is as follows. In section \ref{SnyderReview} we review Snyder's model, following the original presentation. In section \ref{sec:EuclideanSnyder} we use  projective geometry to construct the  Snyder--Euclidean models. Here, we also make a digression to show that different viable momentum spaces can be defined by different choices of  projective coordinates (like for instance Poincar\'e projective coordinates). In this case the phase space algebra takes a different form, although there exists a change of variables relating the different momentum variables. This kind of freedom in the definition of momentum space was briefly mentioned in Snyder's paper, but its meaning can clearly be  understood under the geometric perspective here presented. In section \ref{sec:LorentzianSnyder} we construct the  Snyder--Lorentzian models, showing that we recover Snyder's original model when considering the de Sitter space as the curved momentum space. Here we also show that starting from the Snyder--Lorentzian  models one recovers the  Snyder--Euclidean models when quitting the time and energy variables, a procedure that was until now used to define the ``non-relativistic" Snyder model. 
That this is not the correct procedure becomes clear in section 
\ref{sec:GalileanSnyder}, where we construct the Snyder--Galilean  and the Snyder--Carrollian models by taking, respectively, the  appropriate  $c\to \infty$ and  $c\to 0$ limits of the Snyder--Lorentzian models. The main feature of the Snyder--Galilean models turns out to be the existence of non-vanishing commutation rules between the time and the  space coordinates, and the fact that the associated curved spaces are the ones corresponding to curved Carrollian geometries. In turn, the Snyder--Carrollian models provide a non-vanishing set of commutation rules between time and space coordinates, but now the latter commute among themselves, in contrast to what found in the Galilean case. Finally, the momentum spaces of the Snyder--Carrollian models are provided by the curved analogues of the Galiei spacetimes (the Newton--Hooke spaces). This apparent interchange between the Galilean and Carrollian symmetries of Snyder models is analysed and explained in subsection \ref{GCexplained}. The concluding section includes a schematic table summarizing the main results obtained in the paper and emphasizes the fact that the residual noncommutative interplay between time and space found in the  Galilean and  Carrollian Snyder models is also a recurrent feature of noncommutative Galilean spacetimes with quantum group symmetry, whose role in (2+1) quantum gravity is well-established.


\section{Snyder spacetime revisited}\label{SnyderReview}

In his original work  \cite{Snyder1947} Snyder  investigated whether  spacetime coordinates can be made into noncommuting operators without spoiling special-relativistic Lorentz invariance. While Lorentz invariance might seem in contradiction with the fundamental length scale introduced by the noncommutativity scale,  Snyder made the crucial observation  that the generators of translations in a de Sitter manifold are indeed Lorentz invariant and do not commute among themselves, because of the manifold's curvature. So one can identify these translation generators with actual spacetime coordinates and obtain a discretized spacetime, where coordinates have a Lorentz-invariant spectrum. In this section we briefly revisit Snyder's original work, and we use units such that $\hbar=1$ (in the rest of the paper we will make $\hbar$ explicit). Throughout the paper we use capital Latin letters for indices ranging from 0 to 4, lowercase Latin letters for indices ranging from 1 to 3 and Greek letters for indices ranging from 0 to 3. Our convention for the metric is such that the metric contains mostly minus signs.

The (3+1)-dimensional de Sitter manifold can be constructed via an embedding into a  (4+1)-dimensional Minkowski manifold  defined by the quadratic constraint
\be
-\eta^2=\eta_0^2-\eta_1^2-\eta_2^2-\eta_3^2-\eta_4^2\,,\label{quadratic}
\ee
where $\eta_A$ are the coordinates of the Minkowski manifold and $\eta$ is the de Sitter radius of curvature. The Lorentz group acting on the ambient space is generated by 
\be
M_{AB}=i (\eta_A \partial_B-\eta_B\partial_A),\label{MABrepresentation}
\ee
where  $\partial_A=\eta_{AB}\frac{\partial}{\partial \eta_B}$ and $\eta_{AB}=\text{diag}[1,-1,-1,-1,-1]$. These generators satisfy the commutation relations
\be
\left[M_{AB},M_{CD}\right]=i\left( \eta_{BC}M_{AD}+\eta_{BD}M_{CA}+\eta_{AD}M_{BC}+\eta_{CA}M_{DB} \right)
\ee
and leave the quadratic constraint \eqref{quadratic} invariant.
In particular, the generators of translations $ T_\alpha$, rotations $ L_i$, and boosts $ M_i$ on the de Sitter manifold are:
\be
T_0=  M_{40}\,,  \qquad T_i=M_{4i}\,,\qquad  L_i=\frac12\epsilon_{ijk}  M_{jk}\,,\qquad  M_i= M_{i 0}\,\label{eq:dSsymmetries}
\ee
and satisfy the usual  $\mathfrak{so}(4,1)$ Lie algebra commutation rules:
\be
\begin{array}{lll}
[L_i,L_j]=i\epsilon_{ijk}L_k ,& \qquad [L_i,T_j]=i\epsilon_{ijk}T_k , &\qquad
[L_i,M_j]=i\epsilon_{ijk}M_k , \\[4pt]
\displaystyle{
  [M_i,T_0]=iT_i  } , &\qquad {[M_i,T_j]= i \delta_{ij}  T_0} ,    &\qquad {[M_i,M_j]=- i\epsilon_{ijk} L_k} , 
\\[4pt][T_0,T_i]=-i   M_i , &\qquad   [T_i,T_j]=i  \epsilon_{ijk}L_k , &\qquad[T_0,L_i]=0  .
\end{array}\label{eq:dimensionlessAdS}
\ee

Starting from this de Sitter algebra,  the Snyder model is built by identifying the translation generators with  spacetime coordinates onto a noncommutative Minkowski spacetime:\footnote{Note that physical coordinates are contravariant objects.}
\bea
&& x^i := -a\, T_i =- i a\left( \eta_4 \partial_i- \eta_i  \partial_4\right)=  i a \left( \eta_4 \frac{\partial}{\partial \eta_i}- \eta_i  \frac{\partial}{\partial \eta_4}\right)\,,\nonumber\\
&& x^0 :=\frac a c\, T_0=i \frac a c \left( \eta_4 \partial_0- \eta_0  \partial_4\right)=i \frac a c \left( \eta_4 \frac{\partial}{\partial \eta_0}+ \eta_0  \frac{\partial}{\partial \eta_4}\right)\,, \label{coordinates_representation}
\eea
where we introduced a rescaling by a dimensionful parameter $a$,  with dimensions of length,  and by the speed of light $c$, so to have coordinates with the correct physical dimensions. In this way spacetime  coordinates inherit the noncommutativity of the generators of translations of the  de Sitter algebra, and the parameter $a$ takes the role of a noncommutativity parameter
\be
[x^i,x^j]=a^2 \left(  \eta_i\frac{\partial}{\partial\eta_j}-\eta_j\frac{\partial}{\partial\eta_i} \right),\qquad [x^0, x^i]=- \frac{a^2} c\left(  \eta_i\frac{\partial}{\partial\eta_0}+\eta_0\frac{\partial}{\partial\eta_i} \right)\,.
\ee

It easy to check that   the quadratic form 
\be
S^2=  (x^0)^2-\frac1 {c^2}\Big((x^1)^2+(x^2)^2+(x^3)^2\Big) \,, \label{MinkowskiQuadraticForm}
\ee
constructed with the noncommutative coordinates \eqref{coordinates_representation}, is left invariant by the  generators of rotations and boosts defined in \eqref{eq:dSsymmetries}. So these coordinates are a noncommutative generalization of the Minkowski coordinates and are invariant under the same Lorentz symmetries. 
The Lorentz invariance of the noncommutative spacetime just defined is also reflected in the fact that the commutators between coordinates are covariant under Lorentz transformations. This is made apparent by noticing that the commutators can be written as:
\be
[x^0,x^i]= i\frac{a^2} c M_i\,,\qquad [x^i,x^j]= ia^2\epsilon_{ijk}L_k\,.\label{SnyderSTcommutators}
\ee
Without any need to introduce further structure one can also show that the spatial coordinates just defined have a discrete spectrum  which is Lorentz invariant \cite{Snyder1947}. We stress that the $c$ parameter has been introduced in~\eqref{coordinates_representation} for dimensional reasons. As we will show in the rest of the paper, $c$ must be appropriately included also within the $\mathfrak{so}(4,1)$ algebra and all its associated geometric structures in order to allow us to perform the non-relativistic limit in a consistent way (for instance, the physical boost is $K_i=M_i/c$, see eq.~\eqref{snyderboost}).

The definition of the full Snyder phase space requires   to identify the momenta $p_\alpha$. They are required to  close the usual Poincar\'e algebra with the Lorentz sector (i.e.~behave as vectors under Lorentz transformations) and be  commutative, in analogy to the momenta associated to commutative Minkowski coordinates. So  momenta must satisfy the following relations:\footnote{We remind the reader that physical momenta are covariant objects, as opposed to physical coordinates being contravariant objects, and their physical dimensions are such that $[p_0]=[c\,p_i]$.}
 \bea
\, [p_\alpha,p_\beta] \!\!&=&\!\! 0\,,\nonumber\\
 \,[p_0,M_i]  \!\!&=&\!\! - i  \, c \, p_i\,,\nonumber\\
\,[p_i,M_j]  \!\!&=&\!\! -i\, \delta_{ij}\, \frac{p_0}{c}\,,\\
 \,[p_0,L_i] \!\!&=&\!\! 0\,,\nonumber\\
 \,[p_j,L_i]  \!\!&=&\!\! -i\, \epsilon_{ijk} \, p_k\,.\nonumber
 \eea
 A possible solution  is given by~\cite{Snyder1947}:
 \be
 p_i=\frac{1}{a} \frac{\eta_i}{\eta_4}\,,\qquad p_0=\frac{ c}{a} \frac{\eta_0}{\eta_4}\,,\label{SnyderMomenta}
 \ee
which means that the momenta are proportional to the projective (Beltrami) coordinates of the de Sitter space~\eqref{quadratic}, and therefore the momentum space of the Snyder model is  a de Sitter manifold. The measure on momentum space is thus induced from the de Sitter measure by using \eqref{SnyderMomenta} as the change of variables from the embedding coordinates $\eta_A$ to the  coordinates $p_\alpha$:
 \be
{\rm  d}\tau=\frac{ {\rm  d}p_1 {\rm  d}p_2 {\rm  d}p_3 {\rm  d}p_0}{a  c\left(p_1^2+p_2^2+p_3^2+ \frac1 {a^2 }- \frac{p_0^2} {c^2} \right)^{5/2}}\, .
 \ee 

 In terms of these commutative momenta and the noncommutative coordinates \eqref{coordinates_representation} the Lorentz generators can be represented in the usual way on phase space, 
 \be
 M_i=  c\, p_i x^0 + \frac{1}{c}\, p_0 x^i \,, \qquad L_i=\epsilon_{ijk}x^j p_k\,,
 \label{snyderboost}
 \ee
and here it becomes evident that the physical boost is $K_i=M_i/c$, which has a well-defined  non-relativistic limit $c\to\infty$. 

As a consequence of the previous construction, the commutation rules between the coordinates and momentum operators provide a deformation of the usual Heisenberg--Weyl algebra in terms of the dimensionful parameter $a$, namely:
 \be
 \begin{array}{rcl}
 \left[x^i,p_j\right] \!\!&=&\!\! i \bigl(\delta^i_j+a^2\, p_i p_j \bigr)\,, \\[4pt]
 \left[x^0,p_0\right] \!\!&=&\!\! i \bigl(1-\frac{a^2}{c^2}\, p_0^2 \bigr)\,,\\[4pt]
 \left[x^i,p_0\right] \!\!&=&\!\! i \, a^2\, p_0\, p_i\,,\\[4pt]
 \left[x^0,p_i\right] \!\!&=&\!\! - i \, \frac{a^2}{c^2}\, p_0 \, p_i\,.
 \end{array}
 \label{SnyderModel}
 \ee
The above relations, together with the commutation relations of spacetime coordinates~\eqref{SnyderSTcommutators} and the trivial commutators between momenta, completely characterize the Snyder model in algebraic terms, and give rise to the associated GUP~\cite{Quesne:2006is, Amelino-Camelia:2018wzu, Ivetic:2017gte}. 

In order to consider the physical consequences of the model, this deformed phase space algebra requires a deformed representation of position and momentum operators on a Hilbert space, which is however not qualitatively different from the representation used in standard quantum mechanics. In fact, similarly to what happens in ordinary quantum mechanics, where the position operators can be defined as the generators of translations on a flat momentum space ($\hat x^i =i\partial_{p_i}$) and correspondingly momenta operators are defined as multiplicative operators ($\hat p_i\, \psi(p)=p_i\,\psi(p)$), here the coordinates $x^\alpha$ can still be represented as differential operators acting on functions on the space of momenta (even though they are not just a derivative with respect to momenta) and the momenta $p_\alpha$  act on the same functions in a multiplicative way. This statement will be made more precise in the following section.

To conclude this section, we stress that the lesson learned from Snyder's work is that in order to define noncommutative coordinates without spoiling the classical space(time) symmetries a successful strategy is to identify the coordinates operators with the generators of translations on a curved manifold whose isotropy group describes  the symmetries one wants to preserve. Then the space(time) thus constructed will inherit the symmetries of this manifold. In the following sections we show that one can indeed go beyond Snyder's construction by using different choices of curved  manifolds and using a consistent geometrical prescription for defining the associated momenta.\footnote{A similar construction of a Snyder  spacetime based on the Anti-de Sitter space can be found in the literature~\cite{Mignemi:2011gr} and will be  described in section \ref{sec:LorentzianSnyder}.}


\section{Snyder--Euclidean models} \label{sec:EuclideanSnyder}

As we mentioned, it is possible to replicate the Snyder construction of noncommutative coordinate spaces  and correspondingly deformed phase spaces by starting from different curved manifolds,  constructed as homogeneous spaces whose isotropy group  provides  the symmetries of the associated noncommutative space. As an illustrative simple example, in this section we give a detailed description of this approach for the construction of two-dimensional (2D)  Snyder--Euclidean models, for which the starting curved manifolds can be either the 2D sphere or the hyperbolic plane, endowed with ${\rm SO}(3)$ and ${\rm SO}(2,1)$ symmetry, respectively. As in the original Snyder construction, the spatial coordinates are  identified  with the translation generators on such manifolds, so that the noncommutativity is induced by the curvature of the underlying manifold. The momentum coordinates are again given by the Beltrami projective coordinates on the curved manifold. Together with the space coordinates, they define a deformed phase space.

\subsection{The 2D sphere and the hyperbolic plane}\label{2Dsphere}

Let us consider the one-parametric family of 3D real Lie algebras $\mathfrak{so}_\kk(3)={\rm span}\{P_{1}, P_{2}, J\}$
    with commutation relations given by 
\begin{equation}
 [J ,P_{1}]=   P_2, \qquad  [J ,P_{2}]= -P_1, \qquad [P_{1},P_{2}]=\kk J   , \label{ca}
 \end{equation}
where  $\kk$ is a real parameter.    The Casimir invariant reads
\be
  {\cal C}=P_{1}^2+P_{2}^2+\kk J^2.
 \label{cb}
 \ee
The family $\mathfrak{so}_{\kk}^{(3)}$ comprises    three specific real Lie algebras:   $\mathfrak{so}(3)$ for $\kk>0$,  $\mathfrak{so}(2,1)\simeq \mathfrak{sl}_2(\mathbb R)$ for $\kk<0$, and  $\mathfrak{iso}(2)\equiv \mathfrak{e}(2) = \mathfrak{so}(2) \oplus_S \mathbb{R}^2$ for $\kk=0$, where  hereafter $\oplus_S$ will denote the semidirect sum.  Note that the  value of      $\kk$  can  be 
reduced to $\{+1 , 0 ,-1\}$ through a rescaling of the Lie algebra generators; therefore setting $\kk=0$ in (\ref{ca}) can be shown to be   equivalent to applying  an In\"on\"u--Wigner contraction~\cite{IW}.

The involutive
automorphism  defined by
\be
\Theta  (P_{1},P_{2},J )=(-P_{1},-P_{2},J ) ,
\label{auto}
\ee
generates a $\mathbb Z_2$-grading of   $\mathfrak{so}_\kk^{(3)}$ in such a manner that   $\kk$ is  a
graded contraction parameter~\cite{Montigny}, and  $\Theta$  gives  rise to the following Cartan
decomposition of the Lie algebra:
\be
 \mathfrak{so}_\kk^{(3)} ={\mathfrak{h}}  \oplus  {\mathfrak{p}} ,\qquad 
{\mathfrak{h}  }={\rm span}\{ J  \} =\mathfrak{so}(2) ,\qquad
{\mathfrak{p}  }={\rm span}\{ P_{1},P_{2} \}  .
\label{cxx}
\ee
In this respect, we recall that the usual In\"on\"u--Wigner contraction is applied to a given particular Lie algebra, in our case either $\mathfrak{so}(3)$ with $\kk=+1$ or  $\mathfrak{so}(2,1)$ with $\kk=-1$. The generators anti-invariant under $\Theta$ (\ref{auto}) are multiplied by a dimensionless contraction parameter $\varepsilon$,  while those invariant are kept unchanged, and the limit  $\varepsilon\to 0$ gives rise to the contracted Lie algebra $\mathfrak{iso}(2)$. Clearly this process is equivalent to set $\kk=0$ directly in the commutation relations (\ref{ca}) and, roughly speaking, there is a relationship $\varepsilon\sim \sqrt{\kk}$. Nevertheless, the graded contraction approach goes beyond  In\"on\"u--Wigner contractions, generalizing them, since the former allows one to  also  change the real form as  $\mathfrak{so}(3)$ and  $\mathfrak{so}(2,1)$~\cite{Montigny,MontignyKinematical}; thus, in fact, one deals with a family of real Lie algebras $\mathfrak{so}_\kk(3)$.
Furthermore, as we shall show in the following,  to consider explicitly a graded contraction parameter as $\kk$ (without scaling to $\pm 1$)  from the  beginning in the commutation relations enables us not only to deal simultaneously with simple and non-simple Lie algebras, but to obtain different algebraic and geometric properties in a unified setting (see~\cite{casimirs,conf}).

We denote ${\rm SO}_\kk^{(3)}$ and $H$ the Lie groups with Lie algebras $  \mathfrak{so}_\kk^{(3)} $ and $\mathfrak{h}$ (\ref{cxx}), respectively, and 
we  consider the 2D symmetrical homogeneous
spaces defined by   
\be
\mathbf{S}^2_\kk = {\rm  SO}_{\kk}^{(3)}/H ,\qquad H= {\rm  SO}(2) =\langle J \rangle .
\label{cc}
\ee
These coset spaces  have   constant Gaussian curvature equal to $\kk$ and  are endowed with a  metric with positive  definite signature. 
The generator  $J $ leaves a point $O$ invariant, the origin, so generating  
 rotations around $O$, while $P_{1}$ and $P_{2}$    generate translations which 
 move $O$ along  two basic orthogonal geodesics. Therefore $\mathbf{S}^2_\kk $ (\ref{cc}) describes in a simultaneous way  the three classical 2D Riemannian spaces of constant curvature:
  $$
\begin{array}{lll}
  \mbox{Sphere $\kk>0$:}&\qquad   \mbox{Euclidean plane $\kk=0$:}&\qquad \mbox{Hyperbolic space $\kk<0$:}\\[2pt]
   {\mathbf S}^2={\rm SO}(3)/{\rm  SO}(2)&\qquad    {\mathbf E}^2={\rm  ISO}(2)/{\rm  SO}(2)&\qquad    {\mathbf H}^2= {\rm  SO}(2,1)/{\rm SO}(2)
\end{array}
$$

On these spaces one can define the embedding coordinates $(\s_3,\s_1,\s_2)$, that satisfy the constraint
\begin{equation}
\Sigma_\kk\,  :\ \s_3^2+\kk \bigl( \s_1^2+
  \s_2^2 \bigr)=1 .
\label{ci}
\end{equation}
This provides a ``$\kk$-sphere'', such that for $\kk>0$ we recover the true  sphere, if $\kk<0$ we find the two-sheeted hyperboloid, and in the flat case with $\kk=0$ we get 
 two Euclidean planes $\s_3=\pm 1$ with Cartesian coordinates $(\s_1,\s_2)$.   We   identify the hyperbolic space $\mathbf H^2$  with the connected  component corresponding to the   sheet of the hyperboloid with $\s_3\ge 1$,  and the Euclidean space $\mathbf E^2$   with the plane $\s_3=+1$.

The metric on ${\mathbf
S}^2_\kk$
comes from    the flat ambient metric in $\mathbb R^{3}$ divided by the curvature $\kk$ and
restricted to $\Sigma_\kk$:
\begin{equation}
{\rm d} \sigma_\kk^2=\left.\frac {1}{\kk}
\bigl({\rm d} \s_3^2+   \kk \left( {\rm d} \s_1^2+   {\rm d} \s_2^2\right)
\bigr)\right|_{\Sigma_\kk} \!\! =    \frac{\kk\left(\s_1{\rm d} \s_1 + \s_2{\rm d} \s_2 \right)^2}{1-  \kk \left(\s_1^2+   \s_2^2\right)}+  {\rm d} \s_1^2+   {\rm d} \s_2^2\, .
\label{cj}
\end{equation}
Isometry vector fields in   ambient coordinates for $\mathfrak{so}_{\kk }(3)$, fulfilling     (\ref{ca}),  can be computed and read
\be
P_{1}=\s_3 {\partial_1}   - \kk \, \s_1 \partial_3   ,\qquad
P_{2}=  \s_3 {\partial_2} - \kk  \,\s_2  {\partial_3}    ,\qquad 
J =  \s_2 {\partial_1}    -\s_1 {\partial_2}   ,
\label{ck}
\ee
where $\partial_l=\partial/{\partial \s_l}$.

\subsection{Snyder--Euclidean models}

At this point a noncommutative space compatible with spherical/hyperbolic symmetry can be introduced in full analogy to the original Snyder construction. To this aim, we identify space coordinates  $(x_1,x_2)$ with the   translation generators on the spherical/hyperbolic space:\footnote{Note that  these space coordinates  have the dimensionality of $\sqrt{\omega}$. We will later show that $\sqrt{\omega}$ has dimension of inverse momentum so that upon quantization, $x\to i \hbar\, x$, coordinates turn out to have dimension of length.}
\be
x_1:= P_1\,, \qquad x_2:= P_2\,, \label{EuclideanSnyderCoordinates}
\ee
which implies that they satisfy the following commutator:
\be
\left[x_1,x_2\right]=  \kk J. 
\label{SE2}
\ee

According to the original construction by Snyder, momentum space coordinates are defined by commuting quantities that transform as vectors under the isotropy subgroup $H$, and in this 2D Euclidean case this amounts to impose that they transform as vectors under  rotations. A possible choice, analogous to \eqref{SnyderMomenta}, is to define momenta as 
\be
p_1:= \frac{\s_1}{\s_3}\,,\qquad p_2:= \frac{\s_2}{\s_3}\,.\label{SnyderEuclideanMomenta}
\ee
In the sequel we will show that in fact this choice has a natural interpretation in terms of the projective geometry associated to the geometrical structures behind the noncommutative space construction, since~\eqref{SnyderEuclideanMomenta} correspond to the
so-called Beltrami projective coordinates. However, one could in principle identify momenta with different sets of projective coordinates, such as the Poincar\'e coordinates. This ambiguity is due to the fact that the above-mentioned defining conditions for momenta  do not  identify them uniquely as functions of the ambient coordinates $\s_i$, as was  also remarked in \cite{Snyder1947}.

\subsubsect{Beltrami and Poincar\'e projective coordinates}
 
 The quotients $(\s_1/\s_3,\s_2/\s_3)\equiv (\qq_1,\qq_2)$  of the ambient coordinates (\ref{ci}) are  just the  Beltrami coordinates   of projective geometry for the sphere and the hyperbolic plane~\cite{Doubrovine}. They are obtained by applying   
 the central stereographic projection with pole  
$(0,0,0)\in \mathbb R^{3}$ of a point $ Q=(\s_3, \s_1,\s_2)$ belonging to the projective plane with $\s_3=1$ and coordinates $(\qq_1,\qq_2)$, namely
$$
(\s_3, \s_1,\s_2)=\mu\,
(1,\qq_1,\qq_2),
$$
thus giving rise to the expressions $(i=1,2)$
\begin{equation}
\s_3=\mu=\frac{1}{\sqrt{1+ \kk (\qq_1^2+\qq_2^2)}},\qquad
\s_i=\mu\, \qq_i=\frac{\qq_i}{\sqrt{1+ \kk (\qq_1^2+\qq_2^2)}},  \qquad  \qq_i=\frac{\s_i}{\s_3}  .
\label{da}
\end{equation}
 Thus the   origin  $O=(1,0,0)\in \Sigma_\kk$  projects to the origin $(\qq_1,\qq_2)=(0,0)$ in the projective space ${\mathbf
S}^2_\kk$.  The  domain of $(\qq_1,\qq_2)$   depends on the   value of the curvature $\kk$ since the following condition must be fulfilled
\be
1+ \kk (\qq_1^2+\qq_2^2)>0 .
\label{condition}
\ee
Depending on the sign of $\kk$, we find that:

\begin{itemize}

\item  In the sphere ${\mathbf
S}^2 $ with $\kk>0$,    the condition (\ref{condition}) is automatically satisfied, so that
\be
\qq_i\in(-\infty,+\infty) .
\label{domainS}
\ee
The points in the equator in $ \Sigma_\kk$ with $\s_3=0$ ($\s_1^2+\s_2^2=1/\kk$) go to infinity,  so that the projection (\ref{da}) is   well-defined for the   hemisphere  with $\s_3>0$.

\item
   In the hyperbolic space  ${\mathbf
H}^2$  with $\kk<0$ and    $\s_3\ge 1$, the condition (\ref{condition}) yields
 \be
\qq_1^2+\qq_2^2= \frac{\s_3^2-1}{|\kk|  \s_3^2 }< \frac1{|\kk|} .\label{domainH}
 \ee
 The points at the infinity in $ \Sigma_\kk$ $(\s_3\to\infty)$ are mapped onto  the circle
$ \qq_1^2+\qq_2^2 =  \frac1{|\kk|} $.     

\item
Finally,  in the Euclidean plane ${\mathbf
E}^2$, with $\kk=0$, the Beltrami coordinates are just the Cartesian ones with $\s_i\equiv \qq_i\in(-\infty,+\infty)$ and $\s_3=+1$.

 \end{itemize}

By introducing  (\ref{da}) in the ambient metric (\ref{cj})  we obtain that
\be
{\rm d} \sigma_\kk^2= \frac{(1+\kk\, \>\qq^2) {\rm d} \>\qq^2 - \kk (\>\qq\cdot {\rm d }\>\qq)^2}{(1+\kk\, \>\qq^2)^2} ,
\label{db}
\ee
where $\>\qq=(\qq_1,\qq_2)$ and  hereafter we shall use the following notation for any 2-vectors $\>a=(a_1,a_2)$ and $\>b=(b_1,b_2)$:
\be
\>a^2=a_1^2+a_2^2,\qquad \>a\cdot \>b= a_1 b_1 + a_2 b_2 .
\label{notation}
\ee

A similar construction can be performed by making use of Poincar\'e coordinates $ \bttqq=(\ttqq_1,\ttqq_2)$  which come from 
 the stereographic projection with   pole  $(-1,0,0)\in \mathbb R^{3}$~\cite{Doubrovine}. In this case the relation between the ambient and the projective coordinates reads
 $$
(\s_3 ,\s_1,\s_2)= (-1,0,0)+\lambda\,
(1,\ttqq_1,\ttqq_2),
$$
and we obtain
\be
\lambda=\frac{2}{ {1+ \kk \bttqq^2}},\qquad
\s_3= \lambda -1 =\frac{1- \kk \bttqq^2}{ {1+ \kk \bttqq^2}}   ,\qquad 
\>s=\lambda\, \bttqq=\frac{2\bttqq}{ {1+ \kk \bttqq^2}} ,\qquad    \bttqq=\frac{\>s}{1+\s_3} .
 \label{Poinc}
 \ee
Thus this projection is well defined for any point $Q \in \Sigma_\kk$ except for the   pole $(-1,0,0)$ which projects to the infinity in both the sphere and in the hyperbolic space.  The corresponding domain of Poincar\'e coordinates for each specific space within ${\mathbf
S}^2_\kk$  can be found in~\cite{BaBlHeMu14}. 

Also the Poincar\'e coordinates are  in principle viable momentum space coordinates for the Euclidean Snyder model, since they transform as vectors under rotations and are commutative. In the following subsection we compare the two phase spaces associated, respectively, to the Beltrami and Poincar\'e projective coordinates and discuss their relation.


\subsubsect{Snyder--Euclidean phase  space algebra}

Having defined both noncommutative coordinates and their associate momenta we can now study the properties of the resulting Snyder phase space algebra.
If momenta are identified with Beltrami projective coordinates, direct computations using \eqref{ck}, \eqref{EuclideanSnyderCoordinates} and \eqref{SnyderEuclideanMomenta} lead to the expressions
\be
\begin{array}{ll}
\left[x_1,x_2\right]=  \kk J ,& \quad \left[p_1,p_2\right]= 0,  \\[4pt]
\left[x_1,p_1\right]=1+\kk \, p_1^2,& \quad  \left[x_1,p_2\right]= \kk\, p_1 p_2, \\[4pt]
 \left[x_2,p_2\right]= 1+\kk \, p_2^2, &\quad \left[x_2,p_1\right]= \kk\, p_1 p_2 ,
\end{array}
\label{Snyder2Db}
\ee
where the angular momentum in \eqref{SE2} is represented  as:
\be
 J= x_1\, p_2-x_2\, p_1.
 \label{Jb}
\ee
In order for this model to make sense in  a  quantum-mechanical context we need to define coordinates and momenta as Hermitian operators, $\hat x_i$ and $\hat p_j$, acting on the space of functions $\psi(p)$ defined on the curved momentum space. This is achieved if we define  $\hat x_i:= i\,\hbar\, x_i$ and $\hat p_{j}:=p_{j}$.  In this way we obtain the following deformation of the Heisenberg--Weyl algebra:
\be
\begin{array}{ll}
\left[\hat x_1,\hat x_2\right]= i\hbar\, \kk \hat J ,& \quad \left[\hat p_1,\hat p_2\right]= 0,  \\[4pt]
\left[\hat x_1,\hat p_1\right]=i\hbar\,(1+\kk \, \hat p_1^2),& \quad  \left[\hat x_1,\hat p_2\right]=i\hbar\,  \kk\, \hat p_1 \hat p_2, \\[4pt]
 \left[\hat x_2,\hat p_2\right]= i\hbar\,(1+\kk  \,\hat p_2^2), &\quad \left[\hat x_2,\hat p_1\right]= i\hbar\, \kk \,\hat p_1\hat p_2 ,
\end{array}
\label{Snyder2DbQuantized}
\ee
where  the operators $\hat p_j$ act multiplicatively as usual, $\hat p_j\,\psi(p)=p_j\,\psi(p)$, and the noncommutative coordinates act as $\hat x_i \psi(p)=i\hbar ({\partial_{p_i}}+\kk p_i p_j {\partial_{p_j}}) \psi(p)$.  The quantum angular momentum reads $\hat J= \hat x_1\, \hat p_2-\hat x_2\, \hat p_1$, and it can be easily checked that $J$ and the phase space generators satisfy the Jacobi identities, thus demonstrating that the deformed phase space is indeed invariant under rotational symmetries. Moreover, both $\hat x_{i}$ and $\hat p_{i}$ transform as vector under rotations, as required:
 \be
 [\hat J,\hat x_{1}]= i\hbar \, \hat x_{2},\qquad  [\hat J,\hat x_{2}]=- i\hbar \, \hat x_{1},\qquad  [\hat J,\hat p_1]= i\hbar \, \hat p_2,\qquad  [\hat J,\hat p_2]=- i\hbar \, \hat p_1\,.
 \label{vector}
 \ee

Once the Planck constant has been introduced, we can check that dimensional analysis consistently leads to the correct physical dimensions for space and momentum operators, and that $\omega$ has dimensions of the inverse of squared momentum.

For  
\be
\kk\equiv a^2 >0 
\label{ahbar}
\ee
the phase space algebra~\eqref{Snyder2DbQuantized} exactly recovers the 2D analogue of the space sector of Snyder phase space, see \eqref{SnyderSTcommutators} and \eqref{SnyderModel}. For the sphere $ {\mathbf
S}^2$ with $\kk>0$, we shall say that \eqref{Snyder2DbQuantized} provides the Snyder--Euclidean model $SE^+$, while for $\kk<0$ we have the $SE^-$ model with ${\mathbf
H}^2$ curved momentum space.  
We remark that the momentum space $(p_1,p_2)$ remains commutative for any value of $\kk$, 
 but its domain does depend on $\kk$  according to (\ref{domainS}) and (\ref{domainH}), namely
 \be
\begin{array}{ll}
SE^+:\quad
& p_1,p_2\in(-\infty,+\infty).\nonumber\\
SE^- :\quad  & p_1^{2}+p_2^{2}\in  \bigl[0 ,  1/{|\kk|} \bigr)  .
 \end{array}
\label{domain2D}
\ee

As we mentioned, we could also identify momenta with   the Poincar\'e projective coordinates $\bttqq$. In this case we 
find the  commutation rules
\be
\begin{array}{ll}
\left[ \tilde x_1,\tilde x_2\right]= \kk \tilde J ,& \quad \left[\tilde p_1,\tilde p_2\right]= 0,  \\[4pt]
\left[\tilde x_1,\tilde p_1\right]=\frac 12 \left( 1+\kk (\tilde p_1^2-\tilde p_2^2) \right),& \quad  \left[\tilde x_1,\tilde p_2\right]= \kk\, \tilde p_1 \tilde p_2, \\[4pt]
 \left[\tilde x_2,\tilde p_2\right]=\frac 12 \left( 1+\kk (\tilde p_2^2-\tilde p_1^2) \right), &\quad \left[\tilde x_2,\tilde p_1\right]= \kk\, \tilde p_1 \tilde p_2 ,
\end{array}
\label{Snyder2Dp}
\ee
where the angular momentum  $\tilde J$ is represented as the operator
\be
\tilde J=  2 \, \frac{\tilde x_1\, \tilde p_2-\tilde x_2\, \tilde p_1}{1-\kk (\tilde p_1^2+\tilde p_2^2)}  ,
\label{Jp}
\ee
provided that  $[{\tilde x_1\, \tilde p_2-\tilde x_2\, \tilde p_1},{ \tilde p_1^2+\tilde p_2^2}]=0$.
Hence the expressions (\ref{Snyder2Dp})   provide another way to define  2D Snyder--Euclidean models which are  related to the previous ones (\ref{Snyder2Db}) through the change of variables:
\be
\tilde x_i= x_i,\qquad \tilde p_i=\frac{p_i}{1+\sqrt{1+\kk (p_1^2+p_2^2)}} .
\label{changep}
\ee


\subsubsect{Class of equivalent Snyder--Euclidean phase  space algebras}

The freedom of choosing between different sets of projective coordinates in order to define momenta was already identified in Snyder's original paper \cite{Snyder1947}. There it was noted that the two requirements, that momenta transform as vectors under the relevant group of symmetries and that they are commutative, do not completely fix the functional dependence of  momenta on the  embedding coordinates $(\s_{3},\s_{1},\s_{2})$. 
The   specific  $SE^\pm$ models (\ref{Snyder2Db}) and (\ref{Snyder2Dp}), equivalent through the change of momenta (\ref{changep}),  suggest an alternative way to analyse such a freedom. Instead of paying attention to the embedding coordinates $(\s_{3},\s_{1},\s_{2})$ (\ref{ci}) as the cornerstone to define momenta, one can start from the $SE^\pm$ models already expressed in Beltrami variables given by (\ref{Snyder2Db}) and (\ref{Jb}), so in  the Snyder's choice, and  study the most general change of variables fulfilling the above requirements:

\begin{itemize}

\item The noncommutative Snyder space,  defined by the commutator $\left[x_1,x_2\right]=  \kk J$, is a direct consequence of the Lie algebra of isometries of the underlying curved manifold. 
Since the space coordinates  $x_i$ are identified with the translation generators $P_i$ (\ref{EuclideanSnyderCoordinates}), the former must be 
kept unchanged under any change of variables (as in~(\ref{changep})). This fact conveys that they are transformed as vectors under rotations
due to the commutation relations (\ref{ca}).

 \item Therefore the most general change of variables for  (\ref{Snyder2Db})  only involves momenta and can be expressed as any invertible transformation
 \be
 \tilde p_1=p_1\,f_\kk(p_1,p_2),\qquad  \tilde p_2=p_2\,g_\kk(p_1,p_2),
 \label{transA}
 \ee
such that  $f_\kk$ and $g_\kk$ are smooth functions which also depend on the curvature $\kk$ and  must verify the limits
\be
\lim_{\kk\to 0}f_\kk=a,\qquad \lim_{\kk\to 0}g_\kk=b,
 \label{transB}
\ee
where  $a$ and $b$ are real constants, thus guaranteeing the commutative limit of the $SE^\pm$ models. 

\item Obviously the transformation (\ref{transA}) implies that  the new momenta $ \tilde p_i$ remain commutative, so that it is only necessary to impose that  they behave as vectors under rotations, which leads to two PDEs:
\be
g_\kk= f_\kk+\frac{p_1}{p_2}\left ( p_2\,\frac{\partial f_\kk}{ \partial p_1}-p_1\,\frac{\partial f_\kk}{\partial p_2}\right),\qquad f_\kk= g_\kk+\frac{p_2}{p_1}\left ( p_1\,\frac{\partial g_\kk}{\partial p_2}-p_2\,\frac{\partial g_\kk}{\partial p_1}\right).
 \label{transC}
\ee

\end{itemize}

A straightforward consequence of   this analysis is that the angular momentum $J$ is only represented in its usual form (\ref{Jb}) in the Beltrami variables with $f_\kk\equiv g_\kk=1$, that is, any other possible choice of momenta would give rise to a non-quadratic representation of $J$ as (\ref{Jp}). Therefore the  $SE^\pm$ models in Beltrami variables arise as the distinguished representatives  in the sense that they are the only   phase space algebras defined by quadratic relations.

Notice that a particular subclass of (\ref{transA}) is provided by identifying $f_\kk\equiv g_\kk$ and in this case the  PDEs (\ref{transC}) 
yield
\be
 f_\kk(p_1,p_2)= f_\kk\bigl(p_1^2+p_2^2 \bigr).
\ee
For instance, the $SE^\pm$ models in Poincar\'e variables (\ref{Snyder2Dp}) correspond to the particular choice with
\be
f_\kk\bigl(p_1^2+p_2^2 \bigr)=\frac{1}{1+\sqrt{1+\kk (p_1^2+p_2^2)}} \, ,\qquad \lim_{\kk\to 0}f_\kk=\frac 12.
\ee
 
Indeed,  the above considerations are applicable to any of the generalized Snyder models here presented, but in the rest of this paper we  shall only deal  with Beltrami variables since they are naturally related to the Snyder model in its original (and quadratic) version, and the corresponding phase space algebras for different choices of momenta  can  be  obtained through appropriate change of variables along with the change of the momenta domain. 

One might wonder about  how different choices of projective coordinates  would affect the physical implications of the model. The previous discussion shows that different choices of momenta coordinates  lead to the same kinematical picture, since the Snyder model is uniquely defined by the underlying curved space of momenta, independently of the specific coordinates that are considered onto it. Obviously, the full phase space algebra (from whose $ \left[\hat x_i,\hat p_j\right]$ commutation rules the associated GUP will be derived) is transformed under a given change of momenta coordinates, but note that self-consistency (imposed by the rotational symmetry) is preserved since the noncommutativity between position operators is also changed, being now given by the transformed expression for the rotation generator. Phenomenological differences might arise at the dynamical level, which is however beyond the scopes of the analysis presented here and will be the subject of future work.

Finally, we also remark that the construction of the 3D version of the Snyder--Euclidean model (\ref{Snyder2Db})  is straightforward, and will be presented at the end of section \ref{sec:LorentzianSnyder}.


\section{Snyder--Lorentzian models}\label{sec:LorentzianSnyder}

As we reviewed in section \ref{SnyderReview}, the original (3+1)D Snyder spacetime was constructed by identifying  coordinates with the infinitesimal translation generators of a  (3+1)D de Sitter manifold. In this way spacetime noncommutativity is induced by the curvature of such de Sitter manifold and symmetry under Lorentz transformations is also inherited from the properties of the translation generators of de Sitter as a homogeneous space with Lorentz isotropy subgroup. In this section we generalize the construction to enclose within a unified formalism the two maximally symmetric Lorentzian spaces with constant curvature.
Besides recovering the original Snyder model, we are able to define another Lorentz-invariant noncommutative spacetime starting from the Anti-de Sitter manifold. Also in this case, the momentum space is  identified with the (Beltrami) projective coordinates on the curved manifold.

\subsection{Maximally symmetric Lorentzian spaces}

In order to deal simultaneously with the three Lorentzian manifolds of constant curvature in $(3+1)$ dimensions and to be able to later perform the non-relativistic limit, we introduce a dimensionful redefinition of the generators of the algebra \eqref{eq:dimensionlessAdS}, namely
\be
P_0:=\sqrt{\Lambda} T_0\,,\quad P_i:=\frac{\sqrt{\Lambda}}{c}T_i\,,\quad J_i:=L_i\,,\quad K_i:=\frac 1 c M_i\,.
\label{gallim}
\ee
In this way from  \eqref{eq:dimensionlessAdS} we obtain the two-parameter  family of real Lie algebras that we will denote as $\so=\spn\{ P_0,P_i,K_i,J_i\}$ corresponding, in this order, to the generators of time-like translation, space-like translations, boosts and rotations. Their commutation rules read
\be
\begin{array}{lll}
[J_i,J_j]=\epsilon_{ijk}J_k ,& \qquad [J_i,P_j]=\epsilon_{ijk}P_k , &\qquad
[J_i,K_j]=\epsilon_{ijk}K_k , \\[4pt]
\displaystyle{
  [K_i,P_0]=P_i  } , &\qquad {[K_i,P_j]=\frac 1{c^2}\, \delta_{ij}  P_0} ,    &\qquad {[K_i,K_j]=-\frac 1{c^2}\,\epsilon_{ijk} J_k} , 
\\[4pt][P_0,P_i]=-\ka   K_i , &\qquad   [P_i,P_j]=\ka\, \frac 1{c^2}\, \epsilon_{ijk}J_k , &\qquad[P_0,J_i]=0  ,
\end{array}
\label{fa}
\ee
where  $\ka$ is the cosmological constant that plays the role of a curvature parameter in the Lorentzian manifolds and $c$ is the speed of light. Thus 
$\so$  is isomorphic to   $\mathfrak{so}(4,1)$ for $\ka>0$,  to $\mathfrak{so}(3,2)$ for $\ka<0$,   and  to $\mathfrak{iso}(3,1)=   \mathfrak{so}(3,1) \oplus_S \mathbb{R}^4$ for $\ka=0$.

The family $\so$ of Lie algebras is endowed with two  (second- and fourth-order) Casimir operators~\cite{casimirs}.  The quadratic  one, coming from the Killing--Cartan form, is given by 
\be
{\cal C}
= \frac 1{c^2}\, P_0^2-\>P^2 +\ka \left(   \>K^2-\frac 1{c^2}\, \>J^2\right) ,
\label{fb}
\ee
where   from now on we assume the     notation  (\ref{notation}) for   3-vectors. The fourth-order Casimir can be explicitly found in~\cite{casimirs}.

  The composition of the parity $  \mathcal{P}$  and time-reversal operators $  \mathcal{T}$ defined by~\cite{BLL} 
  \be
   \mathcal{P}( P_0,\>P,\>K,\>J)=  (P_0,-\>P,-\>K,\>J), \qquad
 \mathcal{T}( P_0,\>P,\>K,\>J)=  (-P_0,\>P,-\>K,\>J), 
 \label{ffxx}  \ee
  yields the involutive automorphism
\be
  \mathcal{P}  \mathcal{T}(P_{0},\>P,\>K,\>J )= (-P_{0},-\>P,\>K,\>J ) ,
\label{fc}
\ee
which provides a $\mathbb Z_2$-grading of   $\so$ so that  $\ka$ arises as  a graded contraction parameter~\cite{MontignyKinematical,casimirs}.
The vanishment of $\ka$, which gives the Poincar\'e  algebra, is equivalent to apply  the In\"on\"u--Wigner spacetime contraction~\cite{BLL}, from either $\mathfrak{so}(4,1)$ with $\ka=+1$ or 
$\mathfrak{so}(3,2)$ with $\ka=-1$,   defined by the map $P_\alpha\to  \varepsilon P_\alpha$  and  next  performing the limit $\varepsilon\to 0$, so that $\varepsilon \sim \sqrt{\ka}$. However, as we already commented in section~\ref{2Dsphere} with respect to the family $\mathfrak{so}_\kk(3)$, the explicit presence 
of the parameter $\ka$ allows us to deal with the  three algebras contained within $\so$ and their geometric properties  in a unified way.

 The automorphism (\ref{fc}) leads to the   Cartan
decomposition 
\be
\so ={\mathfrak{h}}  \oplus  {\mathfrak{p}} ,\qquad 
{\mathfrak{h}  }={\rm span}\{\>K, \>J  \} =\mathfrak{so}(3,1) ,\qquad
{\mathfrak{p}  }={\rm span}\{ P_{0},\>P\}  ,
\label{ffcc}
\ee
where $\mathfrak{h}$ and  $\so$ are, in this order, the Lie algebras of  the Lorentz subgroup  $H={\rm SO}(3,1)$   and  the Lie group $\SO$. Hence we obtain a family of $(3+1)$D symmetrical homogeneous    spacetimes 
\be
\mathbf{dS}^{3+1}_\ka = \SO/{\rm SO}(3,1) ,
\label{fe}
\ee
which are of  constant sectional curvature equal to $-\ka$ and whose metric has Lorentzian signature.  According to the sign of $\ka$, we find that $\mathbf{dS}^{3+1}_\ka $ (\ref{fe}) comprises  the  three   (3+1)D  Lorentzian manifolds of constant curvature constructed as homogeneous spaces:
  $$
\begin{array}{lll}
 \mbox{De Sitter $\ka>0$:}&\ \  \mbox{Minkowski $\ka=0$:}&\ \  \mbox{Anti-de Sitter $\ka<0$:}\\[2pt]
   {\bf dS}^{3+1}={\rm SO}(4,1)/{\rm  SO}(3,1)&\ \     {\mathbf M}^{3+1}={\rm  ISO}(3,1)/{\rm  SO}(3,1)&\ \     {\bf AdS}^{3+1}= {\rm  SO}(3,2)/{\rm SO}(3,1)
\end{array}
$$

On these spaces one can define ambient coordinates $(\eta_4,\eta_0,\eta_1,\eta_2,\eta_3)$, which are coordinates on an embedding (4+1)D Minkowski manifold, such that the  $\mathbf{dS}^{3+1}_\ka$ spaces are defined by the constraint \eqref{quadratic}, which we rewrite here for convenience 
\be
\eta_4^2-\eta_0^2+\eta_1^2+\eta_2^2+\eta_3^2=\ka\,.
\ee
Here we have introduced $\ka$ instead of $\eta$ because  in the following we will show that these curved spaces  define the momentum space of the model, so that $\Lambda$ has dimensions of inverse momentum square.
Upon the rescaling
\be
s_0:= \frac{\eta_0}{\Lambda}\,,\quad s_i:=c \frac{\eta_i}{\Lambda}\,,\quad s_4:=\frac{\eta_4}{\sqrt{\Lambda}}\,,
\ee
a new set of ambient coordinates $(\s_4,\s_0,\s_1,\s_2,\s_3)$ can be defined, which satisfy the constraint 
\begin{equation}
\Sigma_\ka\,  :\ \s_4^2-\ka \s_0^2+  \frac{\ka}{c^2}\bigl( \s_1^2+
  \s_2^2+ \s_3^2 \bigr)=1 .
\label{fh}
\end{equation}

Since in the following we will show that these curved spaces  define the geometry of the momentum space of the model, we  take $\Lambda$ to have dimensions of inverse momentum square.
Notice that $\s_4$ is dimensionless but $\s_0$ has dimensions of $1/\sqrt{\ka}$ and $\s_i$ of $c/\sqrt{\ka}$. 
This rescaling was chosen so to make  both the flat $\ka\to 0$ limit and the non-relativistic $c\to \infty$ limit straightforward to perform. In particular, if $\ka=0$ the Minkowski manifold corresponds to the   hyperplane  $\s_4=+1$   with Cartesian coordinates $(\s_0,\>\s)= (\s_0,\s_1,\s_2,\s_3)$. The non-relativistic limit will be discussed in the following section.
The Lorentz subgroup is  just the isotropy subgroup of the point
$O=(1,0,0,0,0)$, that is, the origin in $\mathbf{dS}^{3+1}_\ka $. The  connected component of $\Sigma_\ka$  is identified with the manifold  $\mathbf{dS}^{3+1}_\ka $.

The time-like  metric on $\mathbf{dS}^{3+1}_\ka $ can be obtained   from    the flat ambient metric   divided by the curvature, i.e.~$-\ka$, and
restricted to $\Sigma_\ka$ (\ref{fh}), that is
\begin{equation}
{\rm d} \sigma_\ka^2=\left.\frac {1}{-\ka}
\left({\rm d} \s_4^2-\ka  \dd \s_0^2   +   \frac{\ka}{c^2} \, \dd \>\s^2  
\right)\right|_{\Sigma_\ka} \!\! =    \frac{  -\ka\left(  \s_0{\rm d} \s_0- \frac 1{c^2} \,\>\s \cdot \dd\>\s  \right)^2} {1 + \ka \s_0^2-\frac{\ka}{c^2}\,\>s^2   }+  {\rm d} \s_0^2-\frac 1 {c^2}\,   {\rm d} \>\s^2\, .
\label{fi}
\end{equation}
As expected, when $\ka=0$ we directly recover the  Minkowskian metric: ${\rm d} \sigma^2= {\rm d} \s_0^2-\frac 1 {c^2}\,   {\rm d} \>\s^2$.

Finally, the generators of isometries for the three spaces can be simultaneously written as
\bea
&& P_{0}=\s_4
\,\frac{\partial}{\partial \s_0}+\ka \s_0\, \frac{\partial}{\partial \s_4} \, ,\qquad\ \, P_{i}=-\s_4
\,\frac{\partial}{\partial \s_i}+\frac{\ka}{c^2}\, \s_i\, \frac{\partial}{\partial \s_4} \,  ,  \nonumber\\[4pt]
&&
K_{i}=     \s_0
\,\frac{\partial}{\partial \s_i}+\frac{1}{c^2}\, \s_i\, \frac{\partial}{\partial \s_0} \,  ,\qquad 
J_i=     - \epsilon_{ijk}\, \s_j \, \frac{\partial}{\partial \s_k}  \, ,
\label{fj}
\eea
and it is straightforward to check that they reproduce the Lie algebra~\eqref{fa}.
 

\subsection{Snyder--Lorentzian models from projective geometry}
\label{LSspacetime}

Similarly to the case of the  Snyder--Euclidean models discussed in the previous section,  noncommutative spacetime coordinates that are compatible with Lorentzian symmetry can be defined by identifying them  with the generators of translations on the Lorentzian manifolds discussed in the previous subsection:\footnote{As explained in section \ref{SnyderReview}, the physical spacetime coordinates are contravariant objects.}
\be
x^0:=\frac 1 c\,P_0,\qquad x^i := - c\,P_i\,.\label{LorentzSnyderCoordinates}
\ee 
The factors of $c$ here are introduced so to obtain the correct relation between the dimensionality of the time coordinate and that of the space coordinates, $[c\, x^0]=[x^i]$. Note, however, that these coordinates do not have the dimensions of time and space, respectively, since $[x^0]= [\sqrt{\Lambda}/c]$ and $[x^1]= [\sqrt{\Lambda}]$.  This will be fixed later, once we introduce the constant $\hbar$. In this sense, the model we are constructing only makes sense physically in a quantum-mechanical setting. These coordinates satisfy the following commutation relations:
\be
\left[x^0,x^i\right]=\ka \, K_{i} , \qquad \left[x^i,x^j\right]= \ka\,\epsilon_{ijk} J_k\,,
\ee
inherited from the commutation relations~\eqref{fa} of the symmetry generators of the Lorentzian manifolds.

In the previous section we showed that  the momentum space associated to the noncommutative Euclidean space can be defined by identifying momenta  with the projective coordinates of the Euclidean manifold. This same procedure  also works in the Lorentzian case.
In order to demonstrate this, let us  introduce Beltrami projective coordinates in the Lorentzian manifold $\mathbf{dS}^{3+1}_\ka$. We   apply   
 the  projection with pole  
$(0,0,\>0)\in \mathbb R^{5}$ of a point $ Q=(\s_4, \s_0,\>\s )$ onto the projective hyperplane with $\s_4=1$ and coordinates $(q_0,\>q)=(q_0,q_1,q_2,q_3)$, namely
$$
(\s_4, \s_0,\>\s )=\mu\,
(1,q_0,\>q )\in \Sigma_\ka ,
$$
where
\begin{equation}
\s_4=\mu=\frac{1}{\sqrt{1-\ka q_0^2 +\frac{\ka}{c^2} \>q^2}},\qquad
\s_\alpha=\mu\, q_\alpha=\frac{q_\alpha}{\sqrt{1-\ka q_0^2 +\frac{\ka}{c^2} \>q^2}},  \qquad  q_\alpha=\frac{\s_\alpha}{\s_4}  .
\label{fk}
\end{equation}
Hence this map introduces   coordinates $q_\alpha$  on the half of the   hypersurface $\Sigma_\ka$ (\ref{fh}) with $\s_4>0$.
The   origin  $O=(1,0,\>0)\in \Sigma_\ka$  is mapped onto the origin $(q_0,\>q )=(0,\>0)$ in $\mathbf{dS}^{3+1}_\ka $. 
The projective coordinates  (\ref{fk}) are well-defined whenever
 \be
1-\ka q_0^2 +\frac{\ka}{c^2} \>q^2 >0,
\label{fl}
\ee
so that this condition determines  the admissible   domain of $(q_0,\>q)$, which does depend on the   value of   $\ka$. Only in the   Minkowskian manifold   ${\mathbf
M}^{3+1}$ with $\ka=0$ and given by $\s_4=+1$ the above condition is automatically satisfied and 
 the Beltrami coordinates reduce to the Cartesian ones   $\s_\alpha\equiv q_\alpha\in(-\infty,+\infty)$.   
  
 In these coordinates, the time-like metric in $\mathbf{dS}^{3+1}_\ka $ is directly obtained by   introducing  (\ref{fk}) in   (\ref{fi}): 
\be
{\rm d} \sigma_\ka^2= \frac{\bigl(1- \ka q_0^2 + \frac{\ka}{c^2}   \>q^2  \bigr) \bigl({\rm d} q_0^2 -\frac 1{c^2}{\rm d} \>q^2\bigr) +  \ka \bigr(q_0 \dd q_0- \frac 1 {c^2} \, \>q\cdot {\rm d }\>q \bigl)^2}{ \bigl(1-\ka q_0^2 + \frac{\ka}{c^2}   \>q^2\bigr)^2} .
\label{fm}
\ee
 
Using Beltrami projective coordinates we can define momenta by simply introducing a dimensionful rescaling which allows to recover the correct relation between the dimensionality of energy and that of spatial momenta, $[p_0]=[c\, p_i]$: 
 \be
 p_{0}:= c\,q_0,\qquad p_{i}:= \frac 1 c\, q_i\,.
\label{pa}
\ee
Note that, since $[p_0]=[c\, q_0]=[c\, s_0]=[\frac{ c}{\sqrt{\Lambda}}]$ and $[p_i]=[q_i/c ]=[s_i/c]=[\frac {1}{\sqrt{\Lambda}}]$, they have the correct dimensions of energy and momentum, respectively.
One can easily check that these momenta are commutative and transform as vectors under Lorentz transformations. The condition (\ref{fl}) for the Beltrami coordinates is translated into the following constraint for the momentum space domain:
\be
1+\ka \left( p_{1}^2+p_{2}^2+p_{3}^2  -\frac 1 {c^2}\, p_{0}^2\right)    >0 .
\label{gd}
\ee
Depending on the sign of $\ka$ this implies:
\begin{itemize}

\item  In the de Sitter Lorentzian case  with $\ka>0$,    the condition (\ref{gd}) is  satisfied if $\frac 1 {c^2}\, p_{0}^2-(p_{1}^2+p_{2}^2+p_{3}^2)<\frac 1 \ka$. For on-shell particles this condition implies an upper bound on the mass $c^2 m^2\equiv \frac 1 {c^2}\, p_{0}^2-(p_{1}^2+p_{2}^2+p_{3}^2)$:
\be
c^2  m^2<\frac 1 \ka\,.
\label{domaindS}
\ee

\item
   In the  Anti-de Sitter Lorentzian case  with $\ka<0$ the condition (\ref{gd}) yields
 $p_{1}^2+p_{2}^2+p_{3}^2-\frac 1 {c^2}\, p_{0}^2<\frac 1 {|\ka|}$.
 For on-shell particles this translates into
\be
c^2  m^2>-\frac 1 {|\ka|}\,,
\label{domainAdS}
\ee
which is always satisfied.     

\item
In the Minkowskian case with $\ka=0$, the Beltrami coordinates cover the whole Minkowski manifold and the condition (\ref{gd}) is trivially satisfied.   
 \end{itemize}

Finally, the momentum space metric follows from   \eqref{fm} and \eqref{pa}:
\be
{\rm d} \sigma_{p,\ka}^2= \frac{\bigl(1- \frac{\ka}{c^2} p_0^2 + \ka   \>p^2  \bigr) \bigl( \frac 1{c^2}{\rm d} p_0^2 -{\rm d} \>p^2\bigr) +  \ka \bigr(\frac 1{c^2}p_0 \dd p_0- \>p\cdot {\rm d }\>p \bigl)^2}{ \bigl(1-\frac{\ka}{c^2}p_0^2 +\ka   \>p^2\bigr)^2} .
\label{fm1}
\ee


\subsect{Snyder--Lorentzian phase  space algebra}
\label{42}

 Coordinates \eqref{LorentzSnyderCoordinates} and momenta \eqref{pa} close the following  algebra:
\be
\begin{array}{ll}
\left[x^0,x^i\right]= \ka \, K_{i} ,& \quad \left[x^i,x^j\right]= \ka\,\epsilon_{ijk} J_k,  \\[4pt]
{ \left[x^0,p_{\alpha}\right]=\delta_{0\alpha}-\frac{\ka}{c^2} \, p_{0}p_{\alpha},}& \quad {  \left[x^i,p_{j}\right]=\delta_{ij}+\ka \, p_{i}p_{j}, }
\\[4pt]
\left[x^i,p_{0}\right]=\ka \, p_{0}p_{i},  &\quad  \left[p_{\alpha},p_{\beta}\right]=0,  
\end{array}
\label{gb}
\ee
and in terms of phase space coordinates the Lorentz boosts and rotations  take the usual form
\be
K_{i}= x^0 p_{i}+\frac 1 {c^2} \,  x^i p_{0} ,\qquad J_i= \epsilon_{ijk}  x^{j} p_k.
\ee

As discussed in the previous section, in a quantum-mechanical context coordinates and momenta can be turned into Hermitian operators acting on the space of functions on momenta via the change $\hat x^{\alpha}:= i\hbar\, x^{\alpha}$ and $\hat p_{\alpha}:=p_{\alpha}$. This leads to a generalization of the usual Heisenberg algebra\be
\begin{array}{ll}
\left[\hat x^0,\hat  x^i\right]=  i\hbar\,\ka \, \hat  K_{i} ,& \quad \left[\hat  x^i,\hat  x^j\right]= i\hbar\, \ka\,\epsilon_{ijk} \hat  J_k,  \\[4pt]
{ \left[\hat x^0,\hat p_{\alpha}\right]= i\hbar\,(\delta_{0\alpha}-\frac{\ka}{c^2} \, \hat p_{0}\hat p_{\alpha}),}& \quad {  \left[\hat x^i,\hat p_{j}\right]= i\hbar\,(\delta_{ij}+\ka \, \hat p_{i}\hat p_{j}), }
\\[4pt]
\left[\hat x^i,\hat p_{0}\right]= i\hbar\,\ka \, \hat p_{0}\hat p_{i},  &\quad  \left[\hat p_{\alpha},\hat p_{\beta}\right]=0,
\end{array}
\label{gbQuantum}
\ee
which includes noncommuting spacetime coordinates and nontrivial commutators between space and time sectors of the phase space.  The quantum boost and rotation generators read, respectively,
\be
\hat K_{i}= \hat x^0 \hat p_{i}+\frac 1 {c^2} \,  \hat x^i \hat p_{0} ,\qquad \hat J_i= \epsilon_{ijk}  \hat x^{j} \hat p_k\,,
\ee
and close the following algebra with the phase space coordinates, 
\be
\begin{array}{ll}
    [\hat J_i,\hat x^0]=0, & \quad  [\hat J_i,\hat x^j]= i\hbar \,\epsilon_{ijk}  \hat x^k,\\[4pt]
      [\hat J_i,\hat p_{0}]= 0, &\quad  [\hat J_i,\hat p_{j}]= i\hbar \,\epsilon_{ijk} \hat p_{k} , 
  \end{array}
\label{vectorb}
\ee
\be
\begin{array}{ll}
\displaystyle{     [\hat K_{i},\hat x^0]=- i\hbar \, \frac 1{c^2} \, \hat x^i}, & \quad \displaystyle{   [\hat K_{i},\hat x^j]=- i\hbar \,  \delta_{ij}  \hat x^0},  \\
\displaystyle{   [\hat K_{i},\hat p_{0}]    =  i\hbar \, \hat p_{i} },   &\quad \displaystyle{   [\hat K_{i},\hat p_{j}]= i\hbar \,\frac 1{c^2}\, \delta_{ij} \hat p_{0} }.
 \end{array}
\label{vectorc}
\ee
So  the Jacobi identities are satisfied, thus showing that indeed the  Snyder--Lorentzian phase space algebra  is invariant under Lorentz symmetries. 

Having introduced the physical constants $c$ and $\hbar$ in a consistent way leads to the correct physical dimensions for the spacetime coordinates and for momenta (remember that  we assumed that   $\Lambda$ has the dimension of inverse momentum square, consistently with the interpretation of momenta as the projective coordinates over a curved manifold, with (Anti-)de Sitter geometry). 

The  above expressions in eq.~\eqref{gbQuantum} define what we call the {phase space algebra} of the Snyder--Lorentzian models $SL^\pm$, where the name emphasizes their relativistic invariance properties. 
With $\ka>0$, momenta are the projective coordinates of a de Sitter manifold. In particular, eqs.~\eqref{gbQuantum} reduce to the original Snyder model if
\be
\ka\equiv a^2 >0 \,,
\label{gc}
\ee
up to the constant $\hbar$, which was set to one in section \ref{SnyderReview}. On the other hand, when $\ka<0$ momenta  are the projective coordinates of an Anti-de Sitter manifold (this case was named the Anti-Snyder model in \cite{Mignemi:2011gr}). Note that the momentum operators $\hat p_{\alpha}$ commute for any value of $\ka$, but their allowed values are constrained by the different conditions \eqref{domaindS} and \eqref{domainAdS}.

We remark that the spatial part of (\ref{gbQuantum}) with variables $(\hat x^i,\hat p_{j})$ corresponds to  the commutators
\be
 \left[\hat  x^i,\hat  x^j\right]= i\hbar\, \ka\,\epsilon_{ijk} \hat  J_k,, \qquad  \left[\hat x^i,\hat p_{j}\right]= i\hbar\,(\delta_{ij}+\ka \, \hat p_{i}\hat p_{j}),\qquad 
 \left[\hat p_{i},\hat p_{j}\right]=0,  
 \label{ge}
\ee
which are just the ones for the 3D Snyder--Euclidean models constructed in the previous section, where momenta are the projective coordinates of a sphere when $\ka >0$ and  of a hyperboloid when $\ka <0$, where     $\ka\equiv\kk$ is now the constant sectional curvature of the underlying Riemannian manifold, so generalizing the 2D expressions (\ref{Snyder2DbQuantized}). Therefore we have found the embeddings of the Snyder--Euclidean phase space   within the Snyder--Lorentzian one, and more precisely the embedding of $SE^+$ into $SL^+$ and of  $SE^-$ into $SL^-$.  This restriction of the Snyder--Lorentzian model to the spatial part of the phase space algebra  is what in the literature is called the non-relativistic version of the Snyder model \cite{Mignemi:2011gr, Lu:2011it}. However, in the following we show that  this is not  what one obtains by carefully performing the Galilean limit of the Snyder model.

\section{Snyder--Galilei and Snyder--Carroll  models}\label{sec:GalileanSnyder}

So far we have constructed the Snyder--Euclidean and Snyder--Lorentzian phase space algebras arising, respectively, from the Euclidean and Lorentizan homogeneous spaces with constant curvature. At the level of the phase space algebra, it is rather natural to construct the Snyder--Galilei and the Snyder--Carroll models by taking the appropriate ``non-relativistic" and ``ultra-relativistic" limits of the Snyder--Lorentzian phase space. 
In order to connect these limiting Snyder models to the underlying projective geometry framework, we start from the maximally symmetric spaces associated to the relevant kinematical groups, as done for the Lorentzian case in the previous section.
More precisely, we start by obtaining the Galilei space (and its curved analogues, the Newton--Hooke spaces) by taking the  $c\rightarrow\infty$ of the Lorentzian spaces constructed in the previous section \cite{levy}. The Carroll space (and its  constant curvature analogues) are obtained via the $c\rightarrow 0$ limit of the same Lorentzian spaces \cite{LevyLeblondCarroll}. Note that in order to perform this $c\to0$ limit we first need to  use an appropriate rescaling of the generators of the kinematical group of symmetries, analogue to the one we performed at the beginning of section \ref{sec:LorentzianSnyder}, which was instead optimized to perform the $c\rightarrow\infty$ limit.  In this section we discuss both limits in parallel, since it will turn out that they are  intertwined. 

\subsection{Galilei and Newton--Hooke spaces}

The limit $c\to \infty$ of the Lie algebra $\so$, eq.~\eqref{fa}, can be performed straightforwardly,  giving rise to the family of Newton--Hooke   algebras $\nh$ with Lie brackets and second-order Casimir given by
\be
\begin{array}{lll}
[J_i,J_j]=\epsilon_{ijk}J_k ,& \qquad [J_i,P_j]=\epsilon_{ijk}P_k , &\qquad
[J_i,K_j]=\epsilon_{ijk}K_k , \\[4pt]
\displaystyle{
  [K_i,P_0]=P_i  } , &\qquad {[K_i,P_j]=0} ,    &\qquad {[K_i,K_j]=0} , 
\\[4pt][P_0,P_i]=-\ka   K_i , &\qquad   [P_i,P_j]=0 , &\qquad[P_0,J_i]=0  ,
\end{array}
\label{ha}
\ee
\be
{\cal C}
= -\>P^2 +\ka \, \>K^2  .
\label{hb}
\ee
Notice that this non-relativistic limit is related to the   parity automorphism $  \mathcal{P}$ (\ref{ffxx}) which generates another $\mathbb Z_2$-grading of   $\so$,  such that the factor $1/c^2$ in \eqref{fa} behaves as a graded contraction parameter~\cite{Montigny,MontignyKinematical} (the change of the real forms of the Lie algebras $\so$ corresponds to set $1/c^2<0$, that is, $c\equiv i$).  In terms of the In\"on\"u--Wigner speed-space contraction~\cite{BLL}, one should set $c=1$ in the Lie brackets  \eqref{fa}, map the generators as $\>P\to  \varepsilon \>P$,  $\>K\to  \varepsilon \>K$ and compute the limit $\varepsilon\to 0$; hence   $\varepsilon\sim 1/c$.

We recall that $\nh$ comprises the expanding and oscillating Newton--Hooke algebras as well as the Galilei one~\cite{casimirs,BLL,Aldrovandi,expansiones,pedro,Duval:2014uoa, Figueroa-OFarrill2018} according to the following value of the curvature parameter $\ka$:
\begin{itemize}
\item Expanding Newton--Hooke algebra for  $\ka>0$: 
\be
\begin{array}{l}
\mathfrak{n}^{(3+1)}_+=\mathbb{R}^6 \oplus_S \bigr(  \mathfrak{so}(1,1) \oplus   \mathfrak{so}(3) \bigl) , \\[4pt]
   \mathbb{R}^6=\spn\{\>P,\>K \},\qquad  \mathfrak{so}(1,1)=\spn\{ P_0\},\qquad  \mathfrak{so}(3)= \spn\{\>J \}.
\end{array}
\label{hd}
\ee
 
\item Oscillating Newton--Hooke algebra for  $\ka<0$:
\be
\begin{array}{l}
\mathfrak{n}^{(3+1)}_-= \mathbb{R}^6 \oplus_S \bigr(  \mathfrak{so}(2) \oplus   \mathfrak{so}(3) \bigl)  , \\[4pt]
   \mathbb{R}^6=\spn\{\>P,\>K \},\qquad  \mathfrak{so}(2)=\spn\{ P_0\},\qquad  \mathfrak{so}(3)= \spn\{\>J \}.
\end{array}
\label{he}
\ee

\item Galilei algebra for  $\ka=0$:
\be
\begin{array}{l}
\mathfrak{g}^{(3+1)}\equiv \mathfrak{n}^{(3+1)}_0 = \mathfrak{iiso}(3)= \mathbb{R}^4  \oplus_S  \bigl( \mathbb{R}^3 \oplus_S  \mathfrak{so}(3)  \bigr)  , \\[4pt]
   \mathbb{R}^4=\spn\{P_0,\>P \},\qquad    \mathbb{R}^3=\spn\{ \>K \}, \qquad  \mathfrak{so}(3)= \spn\{\>J \}.
\end{array}
\label{hf}
\ee
\end{itemize}

 The same Cartan decomposition (\ref{ffcc}), associated with the automorphism (\ref{fc}),  also  holds for  $\nh$, but in this case the subalgebra 
$ {\mathfrak{h}  }={\rm span}\{\>K, \>J  \} =\mathfrak{iso}(3)$, that is,  $H={\rm ISO}(3)$ is the Euclidean subgroup of rotations and boosts of the Newton--Hooke and Galilei groups $\NH$. Thus
  we get a family of (3+1)D symmetrical homogeneous  non-relativistic   manifolds  through the coset spaces
   \be
\mathbf{N}^{3+1}_\ka = \NH/{\rm ISO}(3) .
\label{hg}
\ee
Similarly to the Lorentzian manifolds  $\mathbf{dS}^{3+1}_\ka $ (\ref{fe}), the Newtonian manifolds $\mathbf{N}^{3+1}_\ka$  are of  constant sectional curvature equal to $-\ka$, but  the latter are endowed with a degenerate metric (notice that the Killing--Cartan form is degenerate as the Casimir (\ref{hb}) shows). We remark that under the non-relativistic limit $c\to \infty$ (or speed-space contraction~\cite{BLL}),   $\mathbf{dS}^{3+1}_\ka\to  \mathbf{N}^{3+1}_\ka$, so that the de Sitter manifold leads to the expanding Newton--Hooke one (keeping $\ka>0$), the Anti-de Sitter manifold yields   the oscillating Newton--Hooke one (with $\ka<0$), and the Minkowskian manifold gives rise to the flat Galilean space (with $\ka=0$).

We remark that the limit $c\to \infty$ is well-defined in   all the expressions (\ref{fh})--(\ref{fj}).
In particular, in the $c\to\infty$ limit the vector fields \eqref{fj} become
\bea
&& P_{0}=\s_4
\,\frac{\partial}{\partial \s_0}+\ka\, \s_0\, \frac{\partial}{\partial \s_4} \, ,\qquad\ \, P_{i}=-\s_4
\,\frac{\partial}{\partial \s_i} \,  ,  \nonumber\\[4pt]
&&
K_{i}=     \s_0
\,\frac{\partial}{\partial \s_i}\,  ,\qquad 
J_i=     - \epsilon_{ijk}\, \s_j \, \frac{\partial}{\partial \s_k}  \, ,
\label{fj3}
\eea
which fulfil the commutation rules (\ref{ha}) of $\nh$.
 However, the  pseudo-sphere (\ref{fh})  and the time-like metric (\ref{fi}) become degenerate
 \begin{equation}
 \Sigma_\ka\,  :\ \s_4^2-\ka \s_0^2 =1,\qquad {\rm d} \sigma_\ka^2  =  \frac{  {{\rm d}} \s_0^2} {1 + \ka \s_0^2  } \, .
\label{hi}
\end{equation}
This is indeed in agreement with the non-relativistic character of the Newtonian manifolds  $\mathbf{N}^{3+1}_\ka$, since 
the degenerate metric corresponds to an ``absolute-time", here $\s_0$, which generates a foliation (invariant under the corresponding  group action), whose leaves are defined by a constant time $\s_0$. The full metric structure on  $\mathbf{N}^{3+1}_\ka$ is   obtained by considering a ``subsidiary" non-degenerate Euclidean spatial metric restricted to each leaf of the foliation~\cite{conf,pedro} which comes from the limit of  (\ref{fi})  as
\be
{\rm d}{\sigma'}_{\!\!\ka}^2  =\lim_{c\to\infty} \bigl( -c^2 {\rm d} \sigma_\ka^2 \bigr) \quad \mbox{on}\quad \s_0 = {\rm constant},
\ee
yielding
\be
{\rm d}{\sigma'}_{\!\!\ka}^2 =\dd \>\s^2\quad \mbox{on}\quad \s_0 = {\rm constant}.
\label{hij}
\ee

In terms of the Beltrami coordinates (\ref{fk}) the  Newtonian metrics (\ref{hi}) and (\ref{hij}) turn out to be
\be
{\rm d} \sigma_\ka^2  =  \frac{  {{\rm d}} q_0^2} {(1 - \ka q_0^2 )^2 } ,\qquad {\rm d}{\sigma'}_{\!\!\ka}^2 =\frac{\dd \>q^2}{1 - \ka q_0^2 }\quad \mbox{on}\quad q_0 = {\rm constant}.
\ee

\subsection{Carroll space and its curved analogues}
 As we mentioned, the algebra $\so$ as written in eq.~\eqref{fa} is optimized to perform the $c\to\infty$ limit. In order to compute the  $c\to 0$ limit, we  start again from the dimensionless generators of the algebra \eqref{eq:dimensionlessAdS} and perform the following redefinition \cite{LevyLeblondCarroll, Bergshoeff:2014jla, Hartong:2015xda, Cardona:2016ytk, Bergshoeff:2017btm, Gomis:2019, Duval:2014uoa}: 
 \be
  P_0:=c\, \sqrt{\Lambda} \, T_0\,,\quad   P_i:=\sqrt{\Lambda}\, T_i\,,\quad   J_i:=L_i\,,\quad   K_i:=c\, M_i\,.
  \label{carrcontr}
\ee
 Then the $\so$ algebra reads:
 \be
\begin{array}{lll}
[  J_i,   J_j]=\epsilon_{ijk}   J_k ,& \qquad [  J_i,  P_j]=\epsilon_{ijk}  P_k , &\qquad
[  J_i,  K_j]=\epsilon_{ijk}  K_k , \\[4pt]
\displaystyle{
  [  K_i,  P_0]=c^2   P_i } , &\qquad {[  K_i,  P_j]=\delta_{ij}   P_0} ,    &\qquad {[  K_i,  K_j]= - c^2 \epsilon_{ijk}  J_k} , 
\\[4pt][  P_0,  P_i]=-\ka     K_i , &\qquad   [  P_i,   P_j]=\ka \epsilon_{ijk}   J_k , &\qquad[  P_0,  J_i]=0  ,
\end{array}
\label{carr}
\ee
and its Casimir is
\be
{\cal C}=   P_0^2 -c^2 \mathbf{  P}^2+\ka \left( \mathbf{  K}^2-c^2 \mathbf{  J}^2\right) \,.
\ee
 The $c\to0$ limit of this  $\so$ algebra is then straightforward and  gives rise to the family $\mathfrak{c}_\ka^{(3+1)}$ composed by the  Carroll algebra, a para-Euclidean algebra (for $\ka>0$), and a para-Poincar\'e algebra (for $\ka<0$) \cite{BLL} with Lie brackets and second-order Casimir given by
\be
\begin{array}{lll}
[  J_i,   J_j]=\epsilon_{ijk}   J_k ,& \qquad [  J_i,  P_j]=\epsilon_{ijk}  P_k , &\qquad
[  J_i,  K_j]=\epsilon_{ijk}  K_k , \\[4pt]
\displaystyle{
  [  K_i,  P_0]=0 } , &\qquad {[  K_i,  P_j]=\delta_{ij}   P_0} ,    &\qquad {[  K_i,  K_j]=0} , 
\\[4pt][  P_0,  P_i]=-\ka     K_i , &\qquad   [  P_i,   P_j]=\ka \epsilon_{ijk}   J_k , &\qquad[  P_0,  J_i]=0  ,
\end{array}
\label{carrollalgebra}
\ee
\be
{\cal C}
=   P_0^2 +\ka \, \mathbf{  K}^2  .
\label{carrollcasimir}
\ee
We recall that  the limit $c\to0$ of  $\so$ with commutators (\ref{carr}) is  another non-relativistic limit~\cite{LevyLeblondCarroll}, also known as the  ``ultra-relativistic" limit~\cite{Bergshoeff:2014jla,Hartong:2015xda,Cardona:2016ytk,Bergshoeff:2017btm,Gomis:2019}.  This is associated 
 with the   time-reversal automorphism  $ \mathcal{T}$ (\ref{ffxx})
which provides a $\mathbb Z_2$-grading of   $\so$ such that  the factor $c^2$ is now the graded contraction parameter.
The corresponding In\"on\"u--Wigner speed-time contraction~\cite{BLL} requires to set $c=1$ in the commutators  \eqref{carr},  transform the generators as $P_0\to  \varepsilon P_0$,  $\>K\to  \varepsilon \>K$ and perform the limit $\varepsilon\to 0$; thus here  $\varepsilon\sim c$.

The three algebras of the family $\mathfrak{c}_\ka^{(3+1)}$  are characterized as follows:
\begin{itemize}
\item For  $\ka>0$, the para-Euclidean algebra is isomorphic to the Euclidean $\mathfrak{iso}(4)$   algebra but here the former acts as a kinematical algebra:
\be
\begin{array}{l}
\mathfrak{c}_+^{(3+1)} = \mathbb{R}^4 \oplus_S  \mathfrak{so}(4)  ,\quad   \mathbb{R}^4=\spn\{ P_0,\>K\},\quad  
\mathfrak{so}(4)  =\spn\{ \>P,\>J \} .  
\end{array}
\label{c2}
\ee
\item For  $\ka<0$, the para-Poincar\'e algebra so obtained is isomorphic as a Lie algebra to the Poincar\'e algebra $\mathfrak{iso}(3,1)$ but physically different~\cite{BLL}: 
\be
\begin{array}{l}
\mathfrak{c}_-^{(3+1)}\equiv \mathfrak{iso}(3,1) =  \mathbb{R}^4 \oplus_S  \mathfrak{so}(3,1)  ,\quad   \mathbb{R}^4=\spn\{ P_0,\>K\},\quad  
\mathfrak{so}(3,1)  =\spn\{ \>P,\>J \} .  
 \end{array}
\label{c1}
\ee
\item For  $\ka=0$ the Carroll algebra is obtained:
\be
\begin{array}{l}
\mathfrak{c}^{(3+1)}\equiv \mathfrak{c}_0^{(3+1)} \equiv \mathfrak{ii'so}(3) = \mathbb{R}^4 \oplus_S \bigr( \mathbb{R}'^3 \oplus_S  \mathfrak{so}(3)  \bigl) , \\[4pt]
 \mathbb{R}^4=\spn\{ P_0,\>K\},\quad   \mathbb{R}'^3=\spn\{ \>P\},\quad  \mathfrak{so}(3)  =\spn\{\>J \},
\end{array}
\label{c3}
\ee
where this notation means that $ \mathfrak{so}(3) $ acts on  $\mathbb{R}'^3$ through the contragredient of the vector representation and $\mathbb{R}'^3 \oplus_S  \mathfrak{so}(3)$ acts on  $\mathbb{R}^4 $ through the vector representation~\cite{casimirs}. Hence this algebra is not isomorphic to the Galilei one (\ref{hf}) and, in fact, $P_0$ is now a central generator, which provides the Casimir (\ref{carrollcasimir}).
\end{itemize}

 The  Cartan decomposition (\ref{ffcc}), associated with the automorphism (\ref{fc}),    can be applied to  the Carrollian family $\ca$ as well. Similarly to  the Newton--Hooke and Galilean cases, the subalgebra 
$ {\mathfrak{h}  }={\rm span}\{\>K, \>J  \} =\mathfrak{iso}(3)$, that is,  $H$ is the Euclidean subgroup of  the Carrollian group $\rm{C}^{(3+1)}_\ka$. Thus
  we get a family of $(3+1)$D  homogeneous Carrollian manifolds  through the coset spaces
   \be
\mathbf{C}^{3+1}_\ka = \rm{C}_\ka^{(3+1)}/{\rm ISO}(3) .
\label{hg1}
\ee
This means that the  $c\to 0$ limit of the Lorentz subalgebra is $H={\rm ISO}(3)$, which is the common Euclidean subgroup generated by the rotations and boosts of both  the para-Euclidean and para-Poincar\'e groups.

Similarly to   the Lorentzian manifolds  $\mathbf{dS}^{3+1}_\ka $ (\ref{fe}) and the Newtonian manifolds $\mathbf{N}^{3+1}_\ka$ (\ref{hg}), also the Carrollian manifolds $\mathbf{C}^{3+1}_\ka$ (\ref{hg1}) are of  constant sectional curvature, which in this case is equal to $+\ka$. Moreover, the manifolds  (\ref{hg1}) have a degenerate Killing--Cartan form, as the Casimir (\ref{carrollcasimir}) shows, and thus they are endowed with a degenerate metric, just as happens in the Newtonian cases. We remark that under the limit $c\to 0$ (or speed-time contraction~\cite{BLL}),   $\mathbf{dS}^{3+1}_\ka\to  \mathbf{C}^{3+1}_\ka$, so that the de Sitter manifold leads to the positive-curvature Carroll space (which keeps $\ka>0$), the Anti-de Sitter manifold yields   the negative-curvature Carroll space (with $\ka<0$), and the Minkowskian manifold gives rise to the flat Carroll manifold with $\ka=0$. 
  
We remind the reader that the choice of ambient coordinates $\s_A$ in the Lorentzian model discussed in the previous section was dictated by the requirement that the $c\to \infty$ limit is straightforward in those coordinates, as we indeed showed above in this section. However, that choice is not optimal for taking the $c\to 0$ limit. So we perform a different rescaling of the ambient coordinates, namely:
\be
  s_0:= \frac{\eta_0}{c \Lambda}\,,\quad   s_i:= \frac{\eta_i}{\Lambda}\,,\quad   s_4:=\frac{\eta_4}{\sqrt{\Lambda}}\,,
  \label{ambultra}
\ee
such that the constraint (\ref{fh}) reads
\be
 \Sigma_\ka\,  :\  \s_4^2-\ka c^2  \s_0^2+\ka ( \s_1^2+ \s_2^2+ \s_3^2)=1\,.\label{fh2}
\ee
Note that with this rescaling $ \s_4$ is still dimensionless, while $ \s_0$ has dimensions of $1/(c \sqrt\ka)$ and $ \s_i$ of $1/\sqrt\ka$. This different rescaling keeps the correct relation between the dimension of the time and of the space coordinates, $[c  \s_0]=[ \s_i]$.
In terms of these new coordinates the metric  (\ref{fi}) reads
\begin{equation}
{\rm d}  \sigma_\ka^2=\left.\frac {1}{-\ka} 
\left(
{\rm d}  \s_4^2-\ka c^2 \dd  \s_0^2 +   \ka \, \dd \mathbf{ \s}^2\right)
\right|_{\Sigma_\ka}
 \!\! =    \frac{ - \ka\left( c^2  \s_0{\rm d}  \s_0- \mathbf{ \s} \cdot \dd\mathbf{ \s}  \right)^2} {1 + \ka c^2  \s_0^2- \ka\,\mathbf{ \s}^2   }+ c^2 {\rm d}  \s_0^2-   {\rm d} \mathbf{ \s}^2\, ,
\label{fi2}
\end{equation}
which has constant sectional curvature equal to $-\ka$.
When $\ka=0$ we  recover the (rescaled)  Minkowski metric: ${\rm d}  \sigma_{\ka=0}^2= c^2 {\rm d}  \s_0^2-{\rm d} \mathbf{ \s}^2$.
The generators of isometries of this metric are given by a rescaling of the generators \eqref{fj}  written in terms of the new coordinates:
\bea
&&  P_{0} := c P_{0}=  \s_4
\,\frac{\partial}{\partial   \s_0}+\ka c^2   \s_0\, \frac{\partial}{\partial   \s_4} \, ,\qquad\ \,   P_i:= c P_{i}=-  \s_4
\,\frac{\partial}{\partial  \s_i}+\ka\,  \s_i\, \frac{\partial}{\partial  \s_4} \,  ,  \nonumber\\[4pt]
&&
  K_i:= c^2 K_{i}=  c^2    \s_0
\,\frac{\partial}{\partial  \s_i}+  \s_i\, \frac{\partial}{\partial   \s_0} \,  ,\qquad 
  J_i:=J_i=     - \epsilon_{ijk}\,  \s_j \, \frac{\partial}{\partial  \s_k}  \, ,
\label{fj2}
\eea
which fulfill the commutation relations (\ref{carr}).

Now the limit $c\rightarrow 0$ of expressions \eqref{fh2}--\eqref{fj2} is straightforward. The only point to take into account     is that  the limit of the metric (\ref{fi2}) should be $\lim_{c\to 0}( -{\rm d}  \sigma_\ka^2)$,  since   the curvature of the Carrollian manifold (\ref{hg1}) is $+\ka$ instead of $-\ka$.
 In particular,  the  constraint (\ref{fh2})  and the metric (\ref{fi2}) become degenerate
 \begin{equation}
  \Sigma_\ka\,  :\  \s_4^2+\ka ( \s_1^2+ \s_2^2+ \s_3^2)=1,\qquad {\rm d}  \sigma_\ka^2=    \frac{  \ka\left(  \mathbf{ \s} \cdot \dd\mathbf{ \s}  \right)^2} {1 - \ka\,\mathbf{ \s}^2   }+   {\rm d} \mathbf{ \s}^2 \, .
\label{hi1}
\end{equation}

It is worth stressing that the two curved Carrollian manifolds  exactly correspond  to the 3D sphere  ${\mathbf S}^3$ ($\ka>0$) and the hyperbolic space  ${\mathbf H}^3$ ($\ka<0$), as it can be checked from  (\ref{ci}) and (\ref{cj}) for the 2D case, provided that  $\ka\equiv \kk$. This, in turn, means that  under the $c\to 0$ limit the de Sitter manifold ($\ka>0$) leads to the curved Carrollian space $\mathbf{C}^{3+1}_+$ which is endowed with a degenerate  spatial metric determining a curved 3-space of positive constant curvature  (${\mathbf S}^3$). Likewise,  
the Anti-de Sitter manifold ($\ka<0$) gives rise to the curved Carrollian manifold $\mathbf{C}^{3+1}_-$ but  now the degenerate  spatial metric corresponds to a curved 3-space of negative constant curvature (${\mathbf H}^3$). These facts are in agreement with the time- and space-like character of the Lorentzian manifold   $\mathbf{dS}^{3+1}_\ka $. In 
${\mathbf {dS}}^{3+1}$ time-like lines are non-compact  (with generator $P_0$) and space-likes lines are compact (with generators $\>P$) 
so that $\spn\{ \>P,\>J \}=\mathfrak{so}(4) $ as in (\ref{c2}). And, conversely,    in  ${\mathbf {AdS}}^{3+1}$ time-like lines are  compact  while space-likes ones are non-compact, that is, $\spn\{ \>P,\>J \}=\mathfrak{so}(3,1) $ as in (\ref{c1}). These are exactly the subalgebras that remain invariant under the contraction $c\to 0$. 
Therefore, the $c\to 0$ limit is structurally similar to the $c\to \infty$ case discussed above, but with a different role of the time and space sections of the manifold (see also \cite{Gomis:2019}). 
Here, the degenerate metric (\ref{hi1}) corresponds to an ``absolute-space",  which generates a foliation (invariant under the Carrollian  $\rm{C}^{(3+1)}_\ka$ group  action), whose leaves are defined by a constant space, $ \s_i = {\rm constant}$. The full metric structure on  the manifold $\mathbf{C}^{3+1}_\ka$ is   obtained by considering a ``subsidiary" time metric restricted to each leaf of the foliation defined through the limit of~\eqref{fi2} in the form
\be
{\rm d}{  \sigma'}_{\!\!\ka}^2  =
\lim_{c\to0} \left( \frac{1}{c^2}\, {\rm d} \sigma_\ka^2 \right) \quad \mbox{on}\quad \s_i = {\rm constant},
\ee
yielding
\be
{\rm d}{  \sigma'}_{\!\!\ka}^2  ={\rm d} \s_0^2.
\ee

In terms of the Beltrami coordinates on the (Anti-)de Sitter spaces adapted to the  $c\to 0$ limit by making use of \eqref{ambultra}, namely
\begin{equation}
\s_4=\mu=\frac{1}{\sqrt{1-\ka \, c^2 \,q_0^2 +{\ka}\, \>q^2}},\qquad
\s_\alpha=\mu\, q_\alpha=\frac{q_\alpha}{\sqrt{1-\ka \, c^2 \,q_0^2 +{\ka}\, \>q^2}},  \qquad  q_\alpha=\frac{\s_\alpha}{\s_4}  ,
\label{fkc}
\end{equation}
the above two metrics on $ \mathbf{C}^{3+1}_\ka$ turn out to be 
\bea
&&{\rm d} \sigma_\ka^2= \frac{\bigl(1 +  {\ka}  \,  \>q^2  \bigr)  {\rm d} \>q^2 -  \ka \bigr( \>q\cdot {\rm d }\>q \bigl)^2}{ \bigl(1  +  {\ka} \, \>q^2\bigr)^2}   \, , \nonumber\\[4pt]
&&{\rm d}{  \sigma'}_{\!\!\ka}^2    = \frac{\dd q_0^2}{1 + \ka  \>q^2  } \quad\mbox{on}\quad   q_i = {\rm constant}.
\eea

Finally, also the $c\to 0$ limit of the vector fields \eqref{fj2} is straightforward and gives
 \bea
&&  P_{0}=  \s_4
\,\frac{\partial}{\partial   \s_0} \, ,\qquad\ \,   P_i=-  \s_4
\,\frac{\partial}{\partial  \s_i}+\ka\,  \s_i\, \frac{\partial}{\partial  \s_4} \,  ,  \nonumber\\[4pt]
&&
  K_i=    \s_i\, \frac{\partial}{\partial   \s_0} \,  ,\qquad 
  J_i=     - \epsilon_{ijk}\,  \s_j \, \frac{\partial}{\partial  \s_k}  \, .
\label{fj3c}
\eea
It is easy to check that these are compatible with the algebra relations \eqref{carrollalgebra}.

\subsect{Snyder--Galilei and Snyder--Carroll  spacetimes and phase  space algebras}\label{GCexplained}

In this section we work out the Galilei and Carroll limits of the Lorentzian Snyder model that is constructed in section \ref{42}.

As we mentioned in the previous subsection, the coordinates introduced in section  \ref{sec:LorentzianSnyder} are optimized to make the limit $c\to \infty$  well-defined. And indeed we can directly perform such limit  for the deformed Minkowski phase spaces with    brackets (\ref{gb}) and obtain that  
\be
\begin{array}{ll}
\left[x^0,x^i\right]= \ka \, K_{i} ,& \quad \left[x^i,x^j\right]= \ka\,\epsilon_{ijk} J_k,  \\[4pt]
 \left[x^0,p_{\alpha}\right]=\delta_{0\alpha} ,& \quad   \left[x^i,p_{j}\right]=\delta_{ij}+\ka \, p_{i}p_{j}, 
\\[4pt]
\left[x^i,p_{0}\right]=\ka \, p_{0}p_{i},  &\quad  \left[p_{\alpha},p_{\beta}\right]=0,  
\end{array}
\label{ja}
\ee
where now $K_{i}$ are   Galilei boosts and $J_i$ are    rotations,  which read
\be
K_{i}= x^0 p_{i}  ,\qquad J_i=\epsilon_{ijk} x^{j} p_k .\label{KJgalilei}
 \ee

Upon defining the phase space operators $(\hat x^{\alpha}, \hat p_{\alpha})$ we find
\be
\begin{array}{ll}
\left[\hat x^0,\hat  x^i\right]= i\hbar\, \ka \, \hat K_{i} ,& \quad \left[\hat x^i,\hat x^j\right]= i\hbar\,\ka\,\epsilon_{ijk} \hat J_k,  \\[4pt]
 \left[\hat x^0,\hat p_{\alpha}\right]=i\hbar\,\delta_{0\alpha} ,& \quad   \left[\hat x^i,\hat p_{j}\right]=i\hbar\,(\delta_{ij}+\ka \, \hat p_{i}\hat p_{j}), 
\\[4pt]
\left[\hat x^i,\hat p_{0}\right]=i\hbar\, \ka \, \hat p_{0}\hat p_{i},  &\quad  \left[\hat p_{\alpha},\hat p_{\beta}\right]=0.
\end{array}
\label{ja1}
\ee
Consistently, the Galilean boost operators $\hat K_i= \hat x^0 \hat p_{i}$  and the angular momentum operators $\hat J_i=\epsilon_{ijk} \hat x^{j} \hat p_k $ satisfy the following algebra relations:
\be
\begin{array}{ll}
[\hat J_i, \hat x^0]= 0 ,& \quad[\hat J_i,\hat x^j]=i\hbar\,\epsilon_{ijk} \hat x^k,  \\[4pt]
 [\hat J_i,\hat p_{0}]=0 ,& \quad   [\hat J_i,\hat p_{j}]=i\hbar \epsilon_{ijk} \hat p_k\,,
   \end{array}
\ee
\be
\begin{array}{ll}
 [\hat K_i, \hat x^0]= 0 ,& \quad[\hat K_i,\hat x^j]=- i\hbar\,\delta_{ij} \hat x^0,  \\[4pt]
 [\hat K_i,\hat p_{0}]=i\hbar\,\hat p_i ,& \quad   [\hat K_i,\hat p_{j}]=0\,.
\end{array} \label{Galileiinertial}
\ee
It is easy to check that the algebra defined by the phase space operators and the symmetry generators satisfies the Jacobi identities, thus showing that indeed this is a model which is symmetric under Galilean boosts and rotations. Therefore we call this the   {\em Snyder--Galilei phase space algebra}. In particular, the Galilean limit  of the original Snyder model is found by setting 
\be
\ka\equiv a^2>0,
\ee
thus resulting in
\be
\begin{array}{ll}
\left[\hat x^0,\hat x^i\right]= i\hbar\, a^2 \, \hat K_{i} ,& \quad \left[\hat x^i,\hat x^j\right]= i\hbar\,a^2\,\epsilon_{ijk}\hat  J_k,  \\[4pt]
 \left[\hat x^0,\hat p_{\alpha}\right]=i\hbar\,\delta_{0\alpha} ,& \quad   \left[\hat x^i,\hat p_{j}\right]=i\hbar\,(\delta_{ij}+a^2 \, \hat p_{i}\hat p_{j}), 
\\[4pt]
\left[\hat x^i,\hat p_{0}\right]=i\hbar\, a^2 \, \hat p_{0}\hat p_{i},  &\quad  \left[\hat p_{\alpha},\hat p_{\beta}\right]=0.
\end{array}
\label{jaa}
\ee

 Notice that the   spatial part of (\ref{ja1}) generated by $(\hat x^i,\hat p_{i})$  remains invariant under contraction providing the same 
 deformed Euclidean phase spaces (\ref{ge}) as in the original Snyder model. However, in this new model there are nontrivial brackets between the time-like and space-like parts of the phase space, which are instead trivial in the Euclidean models. This difference between the  Snyder--Euclidean and the Snyder--Galilean models is at the root of the  apparent mismatch found in the behaviour of  relativistic and ``non-relativistic'' Snyder particles \cite{Mignemi:2011wh, Mignemi:2013aua, Ivetic:2013yga, Banerjee:2006wf}.
 
 The careful reader will have noticed at this point that, while the $c\to \infty$ limit of the Lorentzian Snyder phase space algebra is well-defined, the connection with the corresponding Newton--Hooke kinematical groups (whose homogeneous spacetimes are the curved counterpart of the Galilean one) seems to be lost. In fact, while in the  Snyder--Galilei phase space both the  time-space and the space-space  commutators for the coordinates are nonzero, in the Newton--Hooke algebra \eqref{ha} the spatial translations commute, so that the Galilei phase space cannot be seen as coming from the identification of the spacetime coordinates with the generators of translations over the space $\mathbf{N}^{3+1}_\ka$.
 Upon further inspection, one surprisingly observes that actually the spacetime commutators in \eqref{ja1} correspond to the commutators of the translation generators of the Carrollian  algebra $\ca$~\eqref{carrollalgebra}. In fact, one can go beyond this observation and show that upon identifying
 \be
\begin{array}{ll}
 x^0:=\,  P_0,&\qquad x^i := -  P_i \\[2pt]
 p_{0}:=   q_0,&\qquad p_{i}:=   q_i\,,
 \end{array}
 \label{ident}
 \ee
 where $  P_\alpha$ are the translations of the $\ca$ algebra, eq.~\eqref{carrollalgebra}, and $  q_\alpha$ are the Beltrami coordinates on the curved Carrollian spaces $\mathbf{C}^{3+1}_\ka$, one recovers exactly the phase space algebra~\eqref{ja} and the symmetry generators \eqref{KJgalilei} are obtained from the corresponding boost and rotation generators in \eqref{fj3}. So the Snyder--Galilei phase space algebra is related to the Carrollian algebra $\ca$, in the sense that the spacetime coordinates are identified with the translation generators over the space $\mathbf{C}^{3+1}_\ka$ and the curved momenta are the Beltrami coordinates on such space.
 
 Conversely, we can also show that the Snyder--Carroll phase space algebra (i.e., the deformed phase space with Carroll symmetries) is actually derived from the Newton--Hooke algebra $\nh$. Let us proceed with the same identification~\eqref{ident}
 where $ P_\alpha$ are now the translations of the $\nh$ algebra (\ref{ha}) and $ q_\alpha$ are the Beltrami coordinates on the Newton--Hooke spaces $\mathbf{N}^{3+1}_\ka$  (\ref{hg}).
 Then one obtains the following phase space commutators:
 \be
\begin{array}{ll}
\left[x^0,x^i\right]= \ka \, K_{i} ,& \quad \left[x^i,x^j\right]= 0\,,  \\[4pt]
 \left[x^0,p_{\alpha}\right]=\delta_{0\alpha} -\ka \, p_0 p_\alpha,& \quad   \left[x^i,p_{j}\right]=\delta_{ij}, 
\\[4pt]
\left[x^i,p_{0}\right]=0,  &\quad  \left[p_{\alpha},p_{\beta}\right]=0,  
\end{array}
\label{ja2}
\ee
where now $K_{i}$ are   Carroll boosts and $J_i$ are rotations, namely
\be
K_{i}= x^i p_{0}  ,\qquad J_i=\epsilon_{ijk} x^{j} p_k ,\label{KJcarroll}
 \ee
which can be derived from the boost and rotation generators in \eqref{fj3c}. Note that, in contrast to the Snyder--Galilean models, the space coordinates of the Snyder--Carrollian models commute among themselves, but still do not commute with the time coordinate.
 
 Upon introducing the phase space operators $(\hat x^{\alpha}, \hat p_{\alpha})$ we find
\be
\begin{array}{ll}
\left[\hat x^0,\hat  x^i\right]= i\hbar\, \ka \, \hat K_{i} ,& \quad \left[\hat x^i,\hat x^j\right]=0 , \\[4pt]
 \left[\hat x^0,\hat p_{\alpha}\right]=i\hbar\,(\delta_{0\alpha} -\ka \, \hat p_{0}\hat p_{\alpha}),& \quad   \left[\hat x^i,\hat p_{j}\right]=i\hbar\,\delta_{ij}, 
\\[4pt]
\left[\hat x^i,\hat p_{0}\right]=0,  &\quad  \left[\hat p_{\alpha},\hat p_{\beta}\right]=0.
\end{array}
\label{ja3}
\ee
Consistently, the Carroll boost operators $\hat K_i= \hat x^i \hat p_{0}$  and the angular momentum operators $\hat J_i=\epsilon_{ijk} \hat x^{j} \hat p_k $ satisfy the following algebra:
\be
\begin{array}{ll}
[\hat J_i, \hat x^0]= 0 ,& \quad[\hat J_i,\hat x^j]=i\hbar\,\epsilon_{ijk} \hat x^k,  \\[4pt]
 [\hat J_i,\hat p_{0}]=0 ,& \quad   [\hat J_i,\hat p_{j}]=i\hbar\,\epsilon_{ijk} \hat p_k\,,
   \end{array}
\ee
\be
\begin{array}{ll}
 [\hat K_i, \hat x^0]= - i\hbar\,\hat x^i ,& \quad[\hat K_i,\hat x^j]=0,  \\[4pt]
 [\hat K_i,\hat p_{0}]=0,& \quad   [\hat K_i,\hat p_{j}]=i\hbar\,\delta_{ij}\hat p_0 \,.
\end{array}\label{Carrollinertial}
\ee
It is easy to check that the phase space coordinates and the symmetry generators satisfy the Jacobi identities (both in the classical and the quantum cases), thus confirming that  this  model is symmetric under the generators of the Carroll boosts and rotations. So we define this a   {\em Snyder--Carroll phase space}. In particular, the Carroll limit  of the original Snyder model is found by setting 
\be
\ka\equiv a^2>0,
\ee
thus resulting in
\be
\begin{array}{ll}
\left[\hat x^0,\hat  x^i\right]= i\hbar\, a^2 \, \hat K_{i} ,& \quad \left[\hat x^i,\hat x^j\right]=0  ,\\[4pt]
 \left[\hat x^0,\hat p_{\alpha}\right]=i\hbar\,(\delta_{0\alpha} -a^2 \, \hat p_{0}\hat p_{\alpha}),& \quad   \left[\hat x^i,\hat p_{j}\right]=i\hbar\,\delta_{ij}, 
\\[4pt]
\left[\hat x^i,\hat p_{0}\right]=0,  &\quad  \left[\hat p_{\alpha},\hat p_{\beta}\right]=0.
\end{array}
\label{ja3b}
\ee

 Notice that, contrary to what happens in the Galilei-invariant and the Lorentz-invariant models, the   spatial part of (\ref{ja3}) with variables $(\hat x^i,\hat p_{i})$  is now trivial, so one recovers a trivial Euclidean  phase space. The nontrivial brackets only appear when  time-like and space-like parts of the phase space are mixed.

 So we have shown that the Galilei-invariant Snyder phase space is derived from the  Carrollian algebras $\ca$, in the sense that the spacetime coordinates are identified with the translation generators over the space $\mathbf{C}^{3+1}_\ka$ and the momenta are just the Beltrami coordinates on such space. Conversely, the Carroll-invariant Snyder phase space  is derived from the  Galilean algebras $\nh$, in the sense that the spacetime coordinates are identified with the translation generators over the space $\mathbf{N}^{3+1}_\ka$ and the momenta are the Beltrami coordinates on such space. 
The reason for this mixing-up of the two limits of the Lorentzian model when going from the kinematical groups to the phase space can be traced back to the fact that in this procedure we are effectively interchanging the role of coordinates and translations, since we identify spacetime coordinates with translations and momenta with Beltrami coordinates. This kind of duality~\footnote{Note that this is not the well-known Galilei-Carroll duality presented in~\cite{Duval:2014uoa}, which refers to the  dual geometric roles played by the Galilean time and the Carrollian time in a Bargmann group context.} is exactly what links the Galilei and Carroll limits of the Lorentz symmetries. To see this, it suffices to compare the action of Galilei and Carroll boosts over noncommutative coordinates and momenta, eqs.~\eqref{Galileiinertial} and \eqref{Carrollinertial}, which we report here for convenience:
\be
\begin{array}{ll}
\text{Galilei boosts}\\[2pt]
 [\hat K_i, \hat x^0]= 0 ,& \quad[\hat K_i,\hat x^j]=- i\hbar\,\delta_{ij} \hat x^0,  \\[2pt]
 [\hat K_i,\hat p_{0}]=i\hbar\,\hat p_i ,& \quad   [\hat K_i,\hat p_{j}]=0\,,
\end{array} 
\quad\begin{array}{ll}
\text{Carroll boosts} \\[2pt]
 [\hat K_i, \hat x^0]= - i\hbar\hat x^i ,& \quad[\hat K_i,\hat x^j]=0,  \\[2pt]
 [\hat K_i,\hat p_{0}]=0,& \quad   [\hat K_i,\hat p_{j}]=i\hbar\,\delta_{ij}\hat p_0 \,,
\end{array}
\label{minitable}
\ee
 (note that this is also the action of Galilei and Carroll boosts over a standard phase space with commutative coordinates and momenta).
 
Equivalently, this interchange between kinematical symmetries can be understood by looking at the Snyder boost generators in their original form~\eqref{snyderboost}, which is optimized to perform the  non-relativistic limit $c\to \infty$ once the physical boosts $K_i=M_i/c$ are considered (see~\eqref{gallim}) 
\be
\lim_{c\to\infty} K_i =\lim_{c\to\infty} M_i/c = \lim_{c\to\infty} {\left(p_i\, x^0 + \frac{1}{c^2} \, p_0\, x^i\right)} = p_i \, x^0,
\ee
thus leading to the Galilean boosts.
On the other hand, the   $c\to 0$ limit can be performed once the physical boosts are defined as $K_i=c\,M_i$ (see~\eqref{carrcontr}) since
\be
\lim_{c\to 0} K_i =\lim_{c\to0} c\, M_i = \lim_{c\to0} {\left(c^2\,p_i\, x^0 +  p_0\, x^i\right)} = p_0\, x^i.
\ee
From these expressions it becomes evident that the flip between Galilean and Carrollian boosts found in this paper is a consequence of interchanging  the roles of space coordinates and momenta. 
This is an essential feature in the Snyder construction which, while being innocuous for the Lorentzian models, has relevant implications in their Galilean and Carrollian counterparts.

\section{Concluding remarks}

In this paper we brought to  a deeper understanding the relationship between the Snyder model for noncommutative spacetime (and the associated deformed phase space) and projective geometry, and uncovered unexpected properties of its Galilean and Carrollian limits.

It was previously understood that Lorentz invariance of the Snyder model is achieved thanks to the fact that spacetime coordinates can be identified with the noncommutative translation generators over an (Anti-)de Sitter manifold and momenta with the Beltrami projective coordinates on the same manifold, so that the curvature parameter of the manifold becomes the spacetime noncommutativity parameter. However, the possibility of using other curved manifolds, invariant under different kinematical groups, in order to construct new noncommutative spacetimes with the corresponding symmetry has never been explored before.

Here we took on this endeavour to explore the Galilean ($c\to\infty$) and Carrollian ($c\to 0$) limits of the Snyder model.
The main  results we obtained  are displayed in  table \ref{tableSummary}. For each group of relativistic symmetries (Lorentzian, Galilean and Carrollian) we constructed the corresponding Snyder phase space algebra (first column of the table), invariant under the rotation and boost generators of the given kinematical group (second column of the table). The Snyder--Galilean and Snyder--Carrollian phase spaces are obtained, respectively, as the $c\to\infty$ and $c\to 0$ limit of the Snyder--Lorentzian one.
Upon analysing the relationship of the  Snyder--Galilean and Snyder--Carrollian models to the projective geometry of curved manifolds, we found that indeed invariance under the Galilean and Carrollian boosts and rotations, respectively, can be ascribed to the underlying curved momentum space structure, upon which coordinates generate translations. 

In particular, the momenta of the Snyder--Galilean phase space are the Beltrami projective coordinates of a curved Carroll manifold (and noncommutative spacetime coordinates are the generators of translations over such a manifold), while the momenta of the Snyder--Carrollian model are the Beltrami projective coordinates of a Newton--Hooke manifold (and noncommutative spacetime coordinates are the generators of translations over such a manifold), see third column of the table. This exchange of the Galilean and Carrollian limits when going from    the phase space picture to the projective geometry picture is due to the exchanged role of coordinates and momenta: coordinates in the projective geometry picture become momenta in the phase space picture and vice-versa.  While the algebra of boosts and rotations is actually the same in the Galilei and Carroll groups of symmetry (compare the entries in the second column of the table), what changes is the action of the generators on phase space. Indeed, the action of the Galilei boosts on coordinates corresponds to the action of Carroll boosts on momenta, and vice-versa (see again the second column of the table and eq.~\eqref{minitable}).

As a byproduct of our investigations, we uncovered previously unknown properties of the non-relativistic Snyder model. This had usually been defined in analogy to the non-relativistic limit of special relativity, where time is absolute, by restricting the Lorentzian model to the space part of the phase space, thereby obtaining a noncommutative space with rotational invariance and an ``absolute time'' which is completely disentangled from space. This corresponds to what we have called the Snyder--Euclidean model. 
We found that this is not the appropriate $c\to \infty$ limit of the Snyder--Lorentzian model. In fact, the noncommutativity parameter (or, equivalently, the momentum space curvature parameter) generates a residual mixing between space and time coordinates in the Galilean limit. This is such that, even though the time coordinate is left invariant by Galilean boosts, still it has a nontrivial commutator with spatial coordinates (a similar, but complementary, effect is found in the Carroll limit, where absolute space has a nontrivial commutator with time).
This residual space-time noncommutativity in the non-relativistic limits is especially relevant in the context of the studies on the Generalized Uncertainty Principle (GUP) \cite{Kempf:1994su, Quesne:2006is, Mignemi:2009ji}, which we mentioned in the introduction. In fact, in light of our results,  applications of the GUP to non-relativistic  physical frameworks cannot simply consider an Euclidean noncommutative spacetime, but must take into account the deformed time-space commutator. We defer to future work the investigation of the implications of this observation to current constraints on GUP and minimal-length scenarios.

In these concluding remarks we would like to place the residual space-time mixing in the non-relativistic limit within a more general quantum gravity perspective. Recall that the Chern--Simons approach to non-relativistic quantum gravity in (2+1) dimensions was developed in \cite{Bernd1,Bernd2} by considering a non-trivial two-fold central extension of the Galilei and Newton--Hooke Lie algebras \cite{BLL,levy} (on the other hand, for an example of deformed Carroll particle emerging from (2+1) gravity see \cite{Kowalski-Glikman:2014paa}). In this context, the time-space noncommutativity also appears in the noncommutative spacetimes obtained when constructing the corresponding  (2+1)D Galilean and Newton--Hooke models. In contrast to the Snyder--Galilean and Snyder--Carrollian models  presented in this paper, within such (2+1)D models with quantum group invariance the time $\hat x^0$ does commute with the noncommutative space coordinates, but nevertheless the commutation rule between space operators does contain the time coordinate. Explicitly, the noncommutative spacetime that arises in these quantum double models is~\cite{pedro}
\be
\left[\hat{x}_1,\hat{x}_2\right]=z\,f_\Lambda\left(\hat{x}_0\right),
\qquad
\left[\hat{x}_1,\hat{x}_0\right]=\left[\hat{x}_2,\hat{x}_0\right]=0,
\label{ncintro}
\ee
where $f_\Lambda\left(\hat{x}_0\right)$ is a formal power series of the `time coordinate' $\hat{x}_0$ whose Galilei (zero curvature) limit $\Lambda\to 0$ is linear in $\hat{x}_0$, hence it does not vanish. Therefore, an ``absolute time" foliation seems to survive, but the model is characterized by an explicit interplay between time and space which is controlled by the Planck scale parameter $z$. Moreover, another well-known noncommutative spacetime model with quantum group invariance is provided by the Galilean limit of the $\kappa$-Minkowski noncommutative spacetime, which reads~\cite{Azcarraga:1995}
\be
\left[\hat{x}^0,\hat{x}^i\right]=-\frac{1}{\kappa}\, \hat{x}^i,
\label{kappaGailiei}
\ee
where $\kappa$ is the Planck mass. This $\kappa$-Galilei spacetime is indeed closer to the models presented in this paper, in the sense that time and space coordinates  no longer commute, although it has to be stressed that in~\eqref{kappaGailiei} the noncommutative spacetime coordinates do generate a subalgebra, whereas this is not the case for the Snyder--Galilean models  in~\eqref{ja1}.


\newpage

\begin{table}[h!]{\footnotesize
\caption{\footnotesize Summary of the  Snyder phase spaces constructed in this paper. There are two models for each possible group of kinematical symmetries (Lorentzian, Galilean and Carrollian), one for each sign of the momentum space curvature parameter $\ka$. For each of these models we report  the deformed phase space algebra (first column), the algebra of transformation generators under which the phase space is invariant, together with the action of the symmetry generators on the phase space (second column) and finally the geometry of the manifold whose Beltrami projective coordinates constitute the momenta of the deformed phase space (third column). Note that the Snyder--Euclidean model is not displayed since it can be straightforwardly inferred by restricting the Snyder--Lorentzian model to the spatial part.
}
\label{tableSummary}
 \begin{center}
\noindent
\begin{tabular}{c c c}
\hline\\[-0.2cm]
\multicolumn{3}{c}{Snyder--Lorentzian model ($SL^\pm$)}\\[0.2cm]
\hline\\
Phase Space Algebra & Phase Space Symmetries&  Momentum Space\\
\\
\multirow{8}{*}{$\displaystyle{\begin{aligned} 
& [\hat x^0\!,\hat  x^i]\!=  i\hbar\,\ka \, \hat  K_{i} &  [\hat  x^i,\hat  x^j]&\!= i\hbar\, \ka\,\epsilon_{ijk} \hat  J_k  \\
&[\hat x^0\!,\hat p_{\alpha}]\!= i\hbar\,\biggl(\delta_{0\alpha}\!-\!\frac{\ka}{c^2} \, \hat p_{0}\hat p_{\alpha}\biggr)&    [\hat x^i,\hat p_{0}]&\!= i\hbar\,\ka \, \hat p_{0}\hat p_{i} \\
&[\hat x^i\!,\hat p_{j}]\!= i\hbar\,\left(\delta_{ij}+\ka \, \hat p_{i}\hat p_{j}\right) &  [\hat p_{\alpha},\hat p_{\beta}]&\!=0
\end{aligned}}$}     &      \multirow{2}{*}{$\displaystyle{\begin{aligned}\, [\hat J_{i},\hat J_{j}]=i \hbar \,\epsilon_{ijk}\hat J_{k}\quad [\hat J_{i},\hat K_{j}]&=i \hbar\,\epsilon_{ijk}\hat K_{k}\\ [\hat K_{i},\hat K_{j}]=- i\hbar\,\frac{1}{c^{2}}  \,\epsilon_{ijk}\hat J_{k} &\end{aligned}}$}      &       \multirow{8}{*}{$\begin{aligned}   \Lambda>0: \;& {\rm SO}(4,1)/{\rm  SO}(3,1)  \\ \Lambda<0:\;&{\rm SO}(3,2)/{\rm  SO}(3,1) \end{aligned}$}\\
 \\
\\
\\
&\multirow{4}{*}{$\begin{aligned}
& [\hat J_i,\hat x^0]=0 &   [\hat J_i,\hat x^j]&= i\hbar \,\epsilon_{ijk}  \hat x^k\\[4pt]
     & [\hat J_i,\hat p_{0}]= 0 &  [\hat J_i,\hat p_{j}]&= i\hbar \,\epsilon_{ijk} \hat p_{k}\\
&   [\hat K_{i},\hat x^0]=- i\hbar \, \frac 1{c^2} \, \hat x^i &    [\hat K_{i},\hat x^j]&=- i\hbar \,  \delta_{ij}  \hat x^0 \\
&  [\hat K_{i},\hat p_{0}]    =  i\hbar \, \hat p_{i}    &    [\hat K_{i},\hat p_{j}]&= i\hbar \,\frac 1{c^2}\, \delta_{ij} \hat p_{0} 
\end{aligned}$}&\\
\\
\\
\\
\\
\\
\\
\\
\hline\\[-0.2cm]
\multicolumn{3}{c}{Snyder--Galilean model ($SG^\pm$)}\\[0.2cm]
\hline\\
Phase Space Algebra & Phase Space Symmetries&  Momentum Space\\
\\
\multirow{8}{*}{$\displaystyle{\begin{aligned} 
& [\hat x^0,\hat  x^i]=  i\hbar\,\ka \, \hat  K_{i} &  [\hat  x^i,\hat  x^j]&= i\hbar\, \ka\,\epsilon_{ijk} \hat  J_k  \\
&[\hat x^0,\hat p_{\alpha}]= i\hbar\,\delta_{0\alpha}&    [\hat x^i,\hat p_{0}]&= i\hbar\,\ka \, \hat p_{0}\hat p_{i} \\
&[\hat x^i,\hat p_{j}]= i\hbar\,(\delta_{ij}\!+\ka \, \hat p_{i}\hat p_{j}) 
 &  [\hat p_{\alpha},\hat p_{\beta}]&=0
\end{aligned}}$}     &      \multirow{2}{*}{$\displaystyle{\begin{aligned}\, [\hat J_{i},\hat J_{j}]=i \hbar \,\epsilon_{ijk}\hat J_{k}\quad [\hat J_{i},\hat K_{j}]&=i \hbar \,\epsilon_{ijk}\hat K_{k}\\ [\hat K_{i},\hat K_{j}]=0&\end{aligned}}$}      &       \multirow{8}{*}{$\begin{aligned}   \Lambda>0: \;& \rm{C}_+^{(3+1)}/{\rm ISO}(3)  \\ \Lambda<0:\;&\rm{C}_-^{(3+1)}/{\rm ISO}(3) \end{aligned}$}\\
 \\
\\
\\
&\multirow{4}{*}{$\begin{aligned}
& [\hat J_i,\hat x^0]=0 &   [\hat J_i,\hat x^j]&= i\hbar \,\epsilon_{ijk}  \hat x^k\\
     & [\hat J_i,\hat p_{0}]= 0 &  [\hat J_i,\hat p_{j}]&= i\hbar \,\epsilon_{ijk} \hat p_{k}\\
&   [\hat K_{i},\hat x^0]=0 &    [\hat K_{i},\hat x^j]&=- i\hbar \,  \delta_{ij}  \hat x^0 \\
&  [\hat K_{i},\hat p_{0}]    =  i\hbar \, \hat p_{i}    &    [\hat K_{i},\hat p_{j}]&=0
\end{aligned}$}&\\
\\
\\
\\
\\
\\
\hline\\[-0.2cm]
\multicolumn{3}{c}{Snyder--Carrollian model ($SC^\pm$)}\\[0.2cm]
\hline\\
Phase Space Algebra & Phase Space Symmetries&  Momentum Space\\
\\
\multirow{8}{*}{$\displaystyle{\begin{aligned} 
&[\hat x^0,\hat  x^i]= i\hbar\, \ka \, \hat K_{i} &  [\hat x^i,\hat x^j]&=0  \\[2pt]
& [\hat x^0,\hat p_{\alpha}]=i\hbar\,(\delta_{0\alpha}\! -\ka \, \hat p_{0}\hat p_{\alpha})&    [\hat x^i,\hat p_{0}]&=0 \\
&[\hat x^i,\hat p_{j}]=i\hbar\,\delta_{ij}  &  [\hat p_{\alpha},\hat p_{\beta}]&=0
\end{aligned}}$}     &      \multirow{2}{*}{$\displaystyle{\begin{aligned}
\, [\hat J_{i},\hat J_{j}]=i \hbar \,\epsilon_{ijk}\hat J_{k}\quad [\hat J_{i},\hat K_{j}]&=i \hbar \,\epsilon_{ijk}\hat K_{k}\\ 
[\hat K_{i},\hat K_{j}]=0 &
\end{aligned}}$}      &       \multirow{8}{*}{$\begin{aligned}   \Lambda>0: \;& \rm{N}_+^{(3+1)}/{\rm ISO}(3) \\ \Lambda<0:\;&\rm{N}_-^{(3+1)}/{\rm ISO}(3) \end{aligned}$}\\
 \\
\\
\\
&\multirow{4}{*}{$\begin{aligned}
& [\hat J_i,\hat x^0]=0 &   [\hat J_i,\hat x^j]&= i\hbar \,\epsilon_{ijk}  \hat x^k\\
     & [\hat J_i,\hat p_{0}]= 0 &  [\hat J_i,\hat p_{j}]&= i\hbar \,\epsilon_{ijk} \hat p_{k}\\
&   [\hat K_{i},\hat x^0]=- i\hbar\,  \hat x^i &    [\hat K_{i},\hat x^j]&=0 \\
&  [\hat K_{i},\hat p_{0}]    = 0    &    [\hat K_{i},\hat p_{j}]&= i\hbar \, \delta_{ij} \hat p_{0} 
\end{aligned}$}&\\
\\
\\
\\
\\
\\
  \hline
\end{tabular}
\end{center}
}
 \end{table}


\section*{Acknowledgments}

This work has been partially supported by Ministerio de Ciencia, Innovaci\'on y Universidades (Spain) under grant MTM2016-79639-P (AEI/FEDER, UE), by Junta de Castilla y Le\'on (Spain) under grants BU229P18 and BU091G19. The authors acknowledge the contribution of the COST Action CA18108.


\end{document}